\documentclass[aps,prb,twocolumn,floatfix]{revtex4}

\usepackage{graphicx}

\begin{document}
\title{Strain controll of magnetic anisotropy in (Ga,Mn)As microbars}
\author{C.~King$^{1}$, J.~Zemen$^{2}$, K.~Olejn\'{\i}k$^{2,3}$, L.~Hor\'ak$^{4}$, J.~Haigh$^{1}$,  V.~Nov\'ak$^{2}$, J.~Ku\v{c}era$^{2}$, V.~Hol\'y$^{4}$, R.~P.~Campion$^{1}$, B.~L.~Gallagher$^{1}$, and T.~Jungwirth$^{2,1}$}
\affiliation{$^{1}$School of Physics and Astronomy, University of Nottingham, Nottingham NG7 2RD, UK}
\affiliation{$^{2}$Institute of Physics ASCR, v. v. i., Cukrovarnick\'a 10, 162 00 Praha 6, Czech Republic}
\affiliation{$^{3}$Hitachi Cambridge Laboratory, Cambridge CB3 0HE, United Kingdom}
\affiliation{$^{4}$Charles University in Prague, Ke Karlovu 3, 121 16 Prague 2, Czech Republic}
\date{\today}

\begin{abstract}
We present an experimental and theoretical study of magnetocrystalline anisotropies in arrays of bars patterned lithographically into (Ga,Mn)As epilayers grown under compressive lattice strain. Structural properties of the (Ga,Mn)As microbars are investigated by high-resolution X-ray diffraction measurements. The experimental data, showing strong strain relaxation effects, are in good agreement with finite element simulations. SQUID magnetization measurements are performed to study the control of  magnetic anisotropy in (Ga,Mn)As by the lithographically induced strain relaxation of the microbars. Microscopic theoretical modeling of the anisotropy is performed based on the  mean-field kinetic-exchange model of the ferromagnetic spin-orbit coupled band structure of (Ga,Mn)As. Based on the overall agreement between experimental data and theoretical modelling we conclude that the micropatterning induced anisotropies are of the magnetocrystalline, spin-orbit coupling origin. 
\end{abstract}
\maketitle

\section{Introduction}
Dilute moment ferromagnetic semiconductors, such as (Ga,Mn)As, are favorable systems for studying and utilizing controllable magnetic anisotropy since micromagnetic parameters of this ferromagnet are very sensitive to  Mn doping, hole concentration, lattice strains, and temperature. The magnetic moment density is small in these ferromagnets and therefore the spin-orbit coupling induced magnetocrystalline anisotropy typically dominates the dipolar-field shape anisotropy. 

The control of the magnetocrystalline anisotropy in (Ga,Mn)As epilayers has been achieved by choosing different substrates and therefore different growth induced strain in the magnetic layer, by varying the growth parameters of the (Ga,Mn)As film, and by postgrowth annealing.\cite{Edmonds:2004_a,Potashnik:2001_a} Reversible electrical control of the magnetocrystalline anisotropy has been demonstrated by utilizing piezo-electric stressors\cite{Rushforth:2008_a,Ranieri:2008_a,Overby:2008_a} or by electrostatic gating in thin-film (Ga,Mn)As field effect transistor structures.\cite{Chiba:2008_a,Owen:2008_a} Recently, a local control of the magnetocrystalline anisotropy has been reported, which provides the possibility for realizing non-uniform magnetization profiles and which can be utilized, e.g., in studies of current induced magnetization dynamics phenomena or non-volatile memory devices.\cite{Wunderlich:2007_c,Pappert:2007_a} In these studies an efficient method of local strain control has been used which is based on lithographic patterning that allows for the relaxation of the lattice mismatch between the (Ga,Mn)As epilayer and the GaAs substrate.\cite{Wenisch:2007_a,Humpfner:2006_a,Wunderlich:2007_c,Pappert:2007_a,Rushforth:2007_a} The modification to the strain distribution  can cause strong changes of the magnetic anisotropy for strains as small as $10^{-4}$.  The high efficiency and practical utility of the lithographic pattering control of magnetic anisotropy in (Ga,Mn)As, demonstrated in the previous works, have motivated our thorough investigation of the phenomenon which is presented in this paper. Our study is based on combined high-resolution X-ray diffraction and magnetization measurements and on macroscopic modeling of the strain relaxation and microscopic calculations of the corresponding magnetic anisotropies. 

We investigate two sets of lithographically patterned (Ga,Mn)As microbars which differ in the thickness to width ratio, Mn doping, and hole concentration. First, we study the structural properties by high resolution X-ray diffraction of microbars patterned in the thicker, higher Mn doped as-grown (Ga,Mn)As material which has a large growth induced strain. The spatial distribution of the lattice relaxation in the stripe cross-section is determined by comparing the measured intensity maps to maps simulated using the theory of elastic deformations and the kinematic scattering theory. The good agreement of the measurement and simulation shows that the applied model is quantitatively reliable in predicting the local lattice relaxation in patterned epilayers subject to small lattice mismatch. This allows us to infer the much weaker lattice relaxation in stripes fabricated in the thinner and lower Mn concentration (Ga,Mn)As by performing only the elastic theory simulations.

In the next step, we measure the magnetic properties of our samples by Superconducting Quantum Interference Device (SQUID) and extract the anisotropy coefficients. Stronger focus is on stripes fabricated in the thinner, annealed (Ga,Mn)As epilayer where the SQUID magnetometry data allow for a reliable extraction of the temperature dependence of the anisotropy coefficients and for direct comparison with the microscopic model. We assumed a linear superposition of the in-plane uniaxial anisotropies and the presence of a single magnetic domain when analyzing the SQUID magnetometry data. We show that the easy axis can be rotated by 90$^{\circ}$ by the micropatterning, completely over-writing the underlying material anisotropy at all studied temperatures.

Finally, we calculate the anisotropy coefficients for a range of material parameters and temperatures below $T_C$. The lattice relaxations determined form the X-ray diffraction measurement  and from finite element simulations  are the inputs of the microscopic calculations of the magnetocrystalline anisotropy. The microscopic model we use  is based on an envelope function description of the valence-band holes and a spin representation for their kinetic-exchange interaction with localized moments on Mn$^{2+}$ ions, treated in the mean-field approximation.\cite{Dietl:2001_b,Abolfath:2001_a,Jungwirth:2006_a,Wenisch:2007_a}

\section{Samples}
\label{se_samples}
\begin{figure}
\includegraphics[scale=0.45]{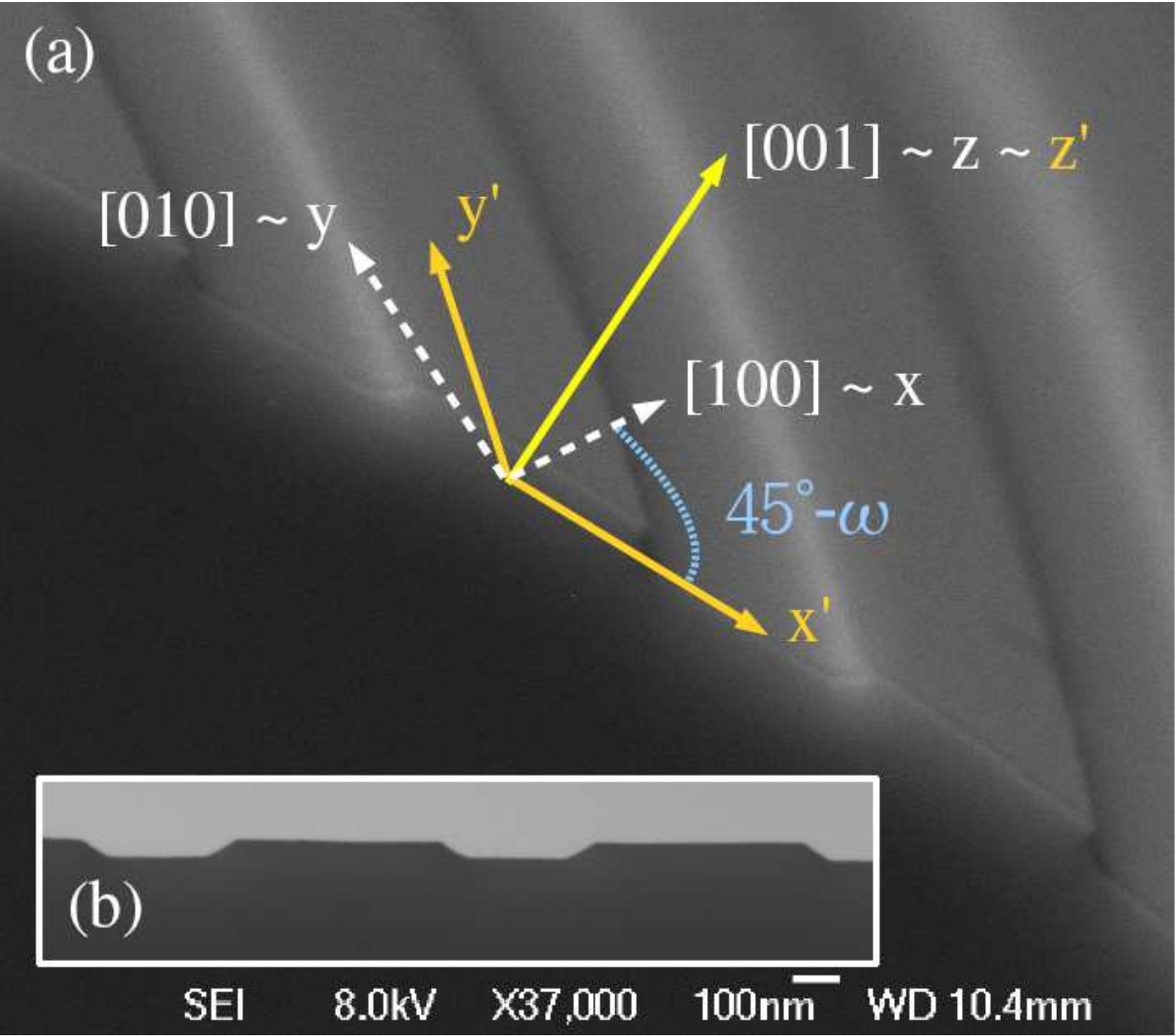}
\caption{(color online) Cross-sectional scanning electron microscope (SEM) images of the stripes in set~A. (a) Image of sample A$_{[110]}$ showing both the cleaved face and the top surface. Although difficult to discern, the profile is undercut. The curvature is due to the sample stage drifting during the exposure of the image. Introduction of coordinates fixed to the crystallographic axes and dashed coordinates fixed to the stripe geometry: the relaxation direction perpendicular to the stripes, the $x'$~axis, is rotated by angle $\omega-45^{\circ}$ with respect to the $[100]$ crystallographic direction, the $x$~axis. The angle $\omega\equiv\alpha-45^{\circ}$ describes the rotation of $x'$ with respect to the $[1\overline{1}0]$ axis. (b) Image of sample A$_{[1\overline{1}0]}$ showing a cut through the stripes and substrate in the $x'-z'$ plane revealing the overcut sides of the stripes.}
\label{sem_Nott}
\end{figure}

We study two sets of patterned (Ga,Mn)As epilayers grown on GaAs substrate. The samples in set~A are doped nominally to $5\%$ of Mn, annealed for approximately 75~minutes at $180^{\circ}$C, and the epilayer is 25~nm thick. The Curie temperature $T_C\approx 120$~K corresponds to optimal annealing of the wafer.\cite{Jungwirth:2005_b} The control sample A$_0$ was not patterned. Samples A$_{[1\overline{1}0]}$ and A$_{[110]}$ were patterned into 25~mm$^2$ arrays of stripes at an angle $\alpha\approx140^{\circ}$ and $\alpha\approx50^{\circ}$, respectively. Here  the angle $\alpha$ is measured from the $[100]$ crystallographic direction. The unintentional $5^{\circ}$ misalignment from the crystal diagonals caused by the microfabrication is accounted for when analyzing the data. The stripes are 750~nm wide, 100~$\mu$m long, and separated by 450~nm gaps, as measured by Atomic Force Microscope (AFM). The fabrication was done by electron beam lithography and wet chemical etching using a solution of phosphoric acid and hydrogen peroxide. The AFM measurements revealed an etch depth of $\approx 60$~nm, and cross-sectional Scanning Electron Microscope (SEM) imaging confirmed that the wet etching leads to anisotropic stripe cross-sections, with the A$_{[110]}$ stripes being undercut and the A$_{[1\overline{1}0]}$ stripes overcut, as shown in Fig.~\ref{sem_Nott}.

The samples in set~B are doped nominally to $7\%$, not annealed, the epilayer is 200~nm thick, and the Curie temperature $T_C\approx 85$~K. The control sample B$_0$ was not patterned. Samples B$_{[1\overline{1}0]}$ and B$_{[010]}$ were patterned into arrays of 1~$\mu$m wide stripes with 1~$\mu$m wide gaps along the $[1\overline{1}0]$ and $[010]$ crystallographic directions, respectively. The fabrication was done by electron beam lithography and dry chemical etching with an etch depth $\approx 700$~nm (B$_{[1\overline{1}0]}$) and $\approx 900$~nm (B$_{[010]}$). The sides of the stripes are slightly overcut in both cases owing to the symmetric dry etching.

With respect to our theoretical modelling of the magnetic anisotropies of our samples, we recall that relating the prediction to the measurement based on the material parameters is not straight forward due to the presence of unintentional compensating defects in the epilayers. Most importantly, a fraction of Mn is incorporated in interstitial positions. These impurities tend to form pairs with Mn$_{\rm Ga}$ acceptors in as-grown systems with approximately zero net moment of the pair, resulting in an effective local-moment concentration $x_{eff}=x_s - x_i$.\cite{Jungwirth:2005_b} Here $x_s$ and $x_i$ are partial concentrations of substitutional and interstitial Mn, respectively. We emphasize that in figures presenting calculated data the Mn concentration labelled as $x$ corresponds to the density of uncompensated local moments, i.e., to $x_{eff}$.

Another input parameter of the theoretical modeling is the lattice mismatch which is different in set~A and~B as it depends on the partial concentrations of Mn atoms in substitutional and interstitial positions in the lattice and of other unintentional impurities.\cite{Masek:2003_a} The lattice mismatch is determined by direct X-ray measurement as detailed in the following section.

Fig.~\ref{sem_Nott} introduces the coordinate system fixed to the crystallographic axes: $x$-axis along the $[100]$ direction, $y$-axis along the $[010]$ direction, and $z$-axis along the $[001]$ direction which is the frame of reference for the microscopic magneto-crystalline anisotropies. The dashed coordinate system is fixed to the stripe geometry: $x'$-axis lies along the relaxation direction transverse to the stripe, $y'$-axis along the stripe, and $z'$-axis along the growth direction coinciding with the $z$-axis. The dashed coordinates are the natural reference for the macroscopic lattice relaxation simulations. 

\section{Lattice relaxation}
\label{se_strain}
The lattice of thin (Ga,Mn)As films grown epitaxially on GaAs substrates is strained compressively due to a lattice mismatch $e_0 = (a_s-a_0)/a_0 < 0$ where $a_s$ and $a_0$ are the lattice constant of the substrate and of the relaxed free-standing (Ga,Mn)As epilayer, respectively. The narrow stripes allow for anisotropic relaxation of the compressive strain present in the unpatterned epilayer. An expansion of the crystal lattice along the direction perpendicular to the bar occurs while the epilayer lattice constant along the bar remains unchanged. Parameters sufficient for determination of the induced strain are the lattice mismatch $e_0$ and the shape of the stripe, mainly the thickness to width ratio of the stripe. In the regime of small deformations the components of the induced strain are linearly proportional to the lattice mismatch. The strain tensor in the coordinate system fixed to the stripe reads:
\begin{eqnarray}
\label{strain_relax_comsol} 
{\bf e}^{r} & = & e_0\left(\begin{array}{ccc} 
-\rho + 1 & 0 & 0 \\
0 & 1 & 0 \\ 
0 & 0 & \frac{c_{12}}{c_{11}}\left(\rho - 2\right) \\
\end{array}\right), \;
\end{eqnarray}
where the lattice relaxation is quantified by $\rho(x',z')$ which varies over the stripe cross-section, $c_{12}$ and $c_{11}$ are the elastic moduli. The strain components in this work are expressed with respect to a relaxed free-standing (Ga,Mn)As epilayer. Note that for $\rho = 0$ the strain tensor ${\bf e}^{r}$ describes the growth strain of the unpatterned epilayer.
In this section we investigate experimentally and theoretically the geometry of the stripes, the size of the lattice mismatch and the spatial dependence of the lattice relaxation $\rho(x',z')$. The results are used as an input of the microscopic modeling of the magnetic anisotropies in Sec.~\ref{se_theory}.

Microbars in set~B have larger thickness to width ratio than microbars in set~A. Therefore the relaxation is expected to be larger in set~B. At the same time, the (Ga,Mn)As epilayer has larger volume in set~B, primarily due to a larger number of interstitial Mn in this higher doped unannealed material. The larger film thickness and larger growth strain in set~B make these materials significantly more favorable for an accurate X-ray diffraction analysis of the strain profile in the patterned microbars.

The lattice relaxation in samples B$_{[1\overline{1}0]}$ and B$_{[110]}$ was measured by high-resolution X-ray diffraction using the synchrotron source at ESRF Grenoble (beamline ID10B, photon energy 7.95~keV). For a reliable determination of both in-plane ($u'_x$) and vertical ($u'_z$) components of the elastic displacement field we measured the reciprocal-space distribution of the diffracted intensity around the symmetric 004 and asymmetric 404 reciprocal lattice points. The asymmetric diffraction was chosen so that the in-plane component of the corresponding reciprocal lattice vector $h$ was perpendicular to the stripes. The diffracted radiation was measured by a linear X-ray detector lying in the scattering plane.

\begin{figure}
\includegraphics[scale=0.45]{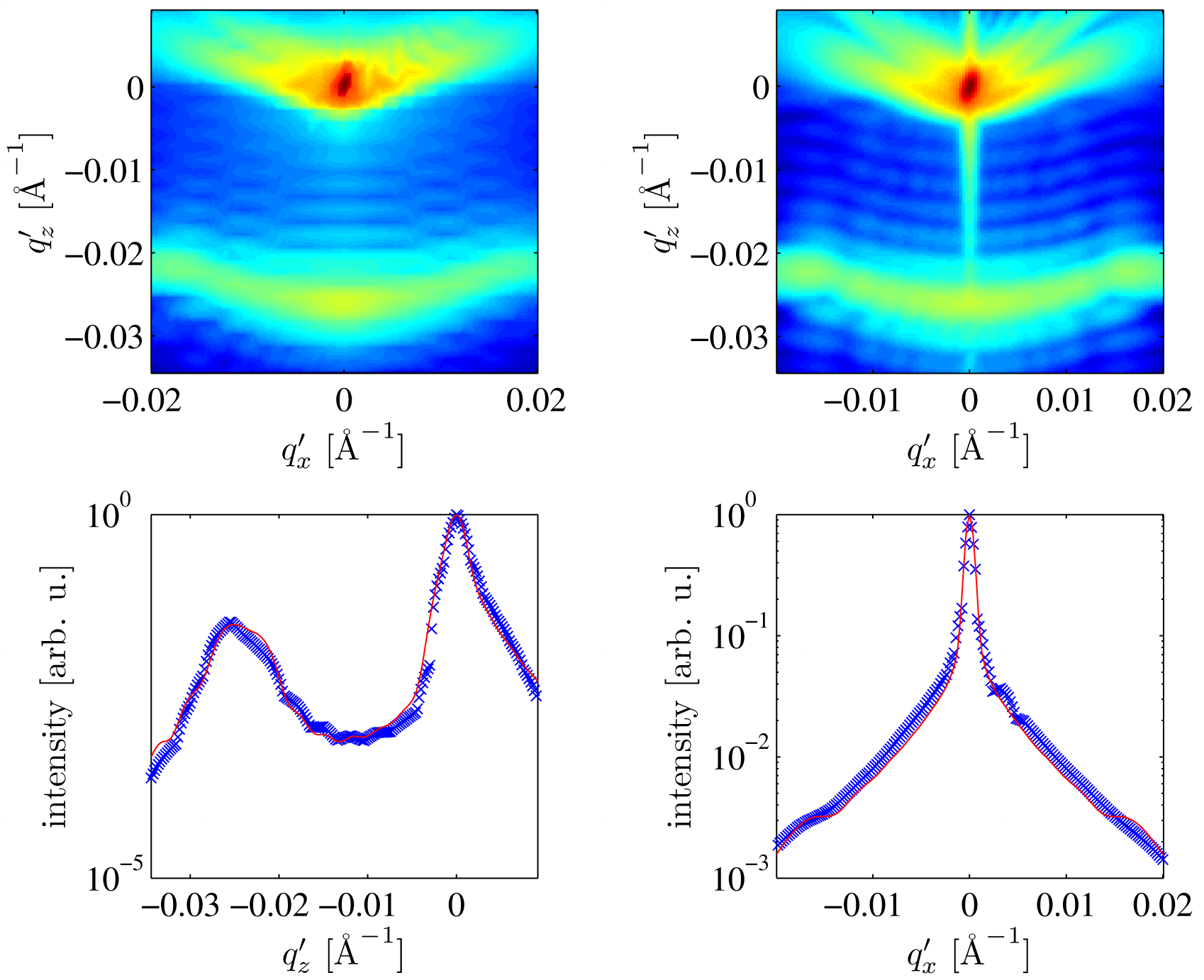}
\caption{(Color online) The measured (upper left panel) and simulated (upper right panel) reciprocal-space maps in the symmetric 004 diffraction of sample B$_{[010]}$. In the bottom row, the measured (points) and simulated (lines) intensities integrated along the horizontal (left) and vertical (right) directions are plotted. In the intensity maps, the color scale is logarithmic.}
\label{diffract_004}
\end{figure}

\begin{figure}
\includegraphics[scale=0.45]{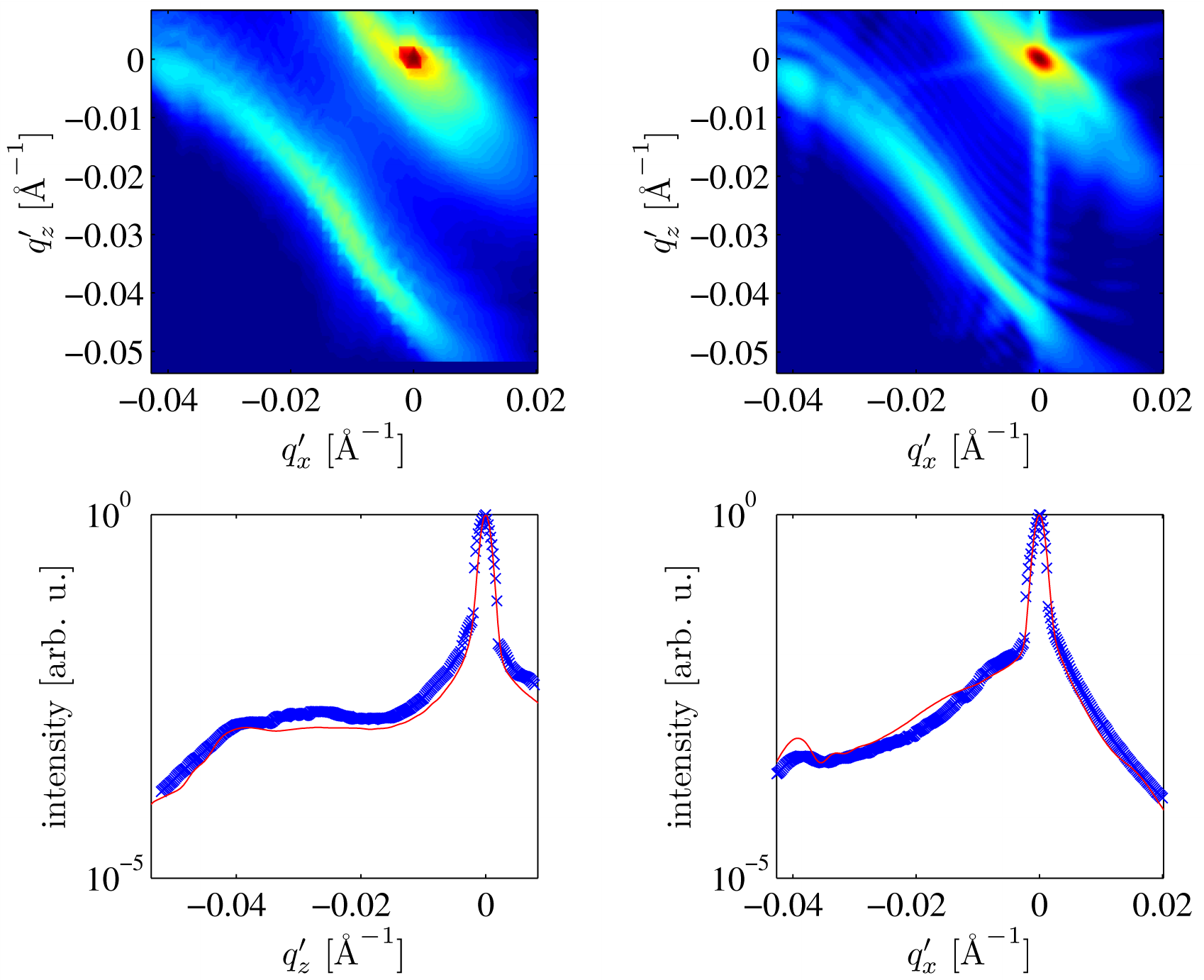}
\caption{(Color online) The measured (upper left panel) and simulated (upper right panel) reciprocal-space maps in the asymmetric 404 diffraction of sample B$_{[010]}$. In the bottom row, the measured (points) and simulated (lines) intensities integrated along the horizontal (left) and vertical (right) directions are plotted. In the intensity maps, the color scale is logarithmic.}
\label{diffract_404}
\end{figure}

Figs.~\ref{diffract_004} and~\ref{diffract_404} present examples of the measured (upper left panels) and simulated (upper right panels) reciprocal space maps, showing two maxima corresponding to the reciprocal lattice points of the GaAs substrate and the (Ga,Mn)As layer. The bottom panels show the measured and simulated integrated intensities for two directions in the reciprocal space. Since the lateral stripe period was larger than the coherence width of the primary radiation, different stripes were irradiated incoherently, so that the lateral intensity satellites stemming from the lateral stripe periodicity could not be resolved. The measured intensity distribution is therefore proportional to the intensity scattered from a single microbar. 

We fitted the measured intensity maps to numerical simulations based on the kinematic scattering theory and the theory of anisotropic elastic medium. We used a finite-element simulation (implemented in Structural Mechanics Module of Comsol Multiphysics, standard partial differential equation solver) to obtain the local relaxation distribution $\rho(x',z')$ in the stripes and derived the corresponding reciprocal space map. The angle of the sides of the stripes and the lattice mismatch $e_0$ of the (Ga,Mn)As and GaAs lattices were the two fitting parameters. The left column of Figs.~\ref{diffract_004} and~\ref{diffract_404} shows the measured diffraction maps and projections. The right column shows the simulated results. The lateral and vertical projections of the measured and simulated intensity maps as well as the whole maps are used in the fitting. The coordinates $q'_x$ and $q'_z$ span the reciprocal space conjugate to the real space with coordinates $x'$ and $z'$ fixed to the stripe. They are measured with respect to the reciprocal lattice point 004 and 404.

\begin{figure}
\includegraphics[scale=0.3]{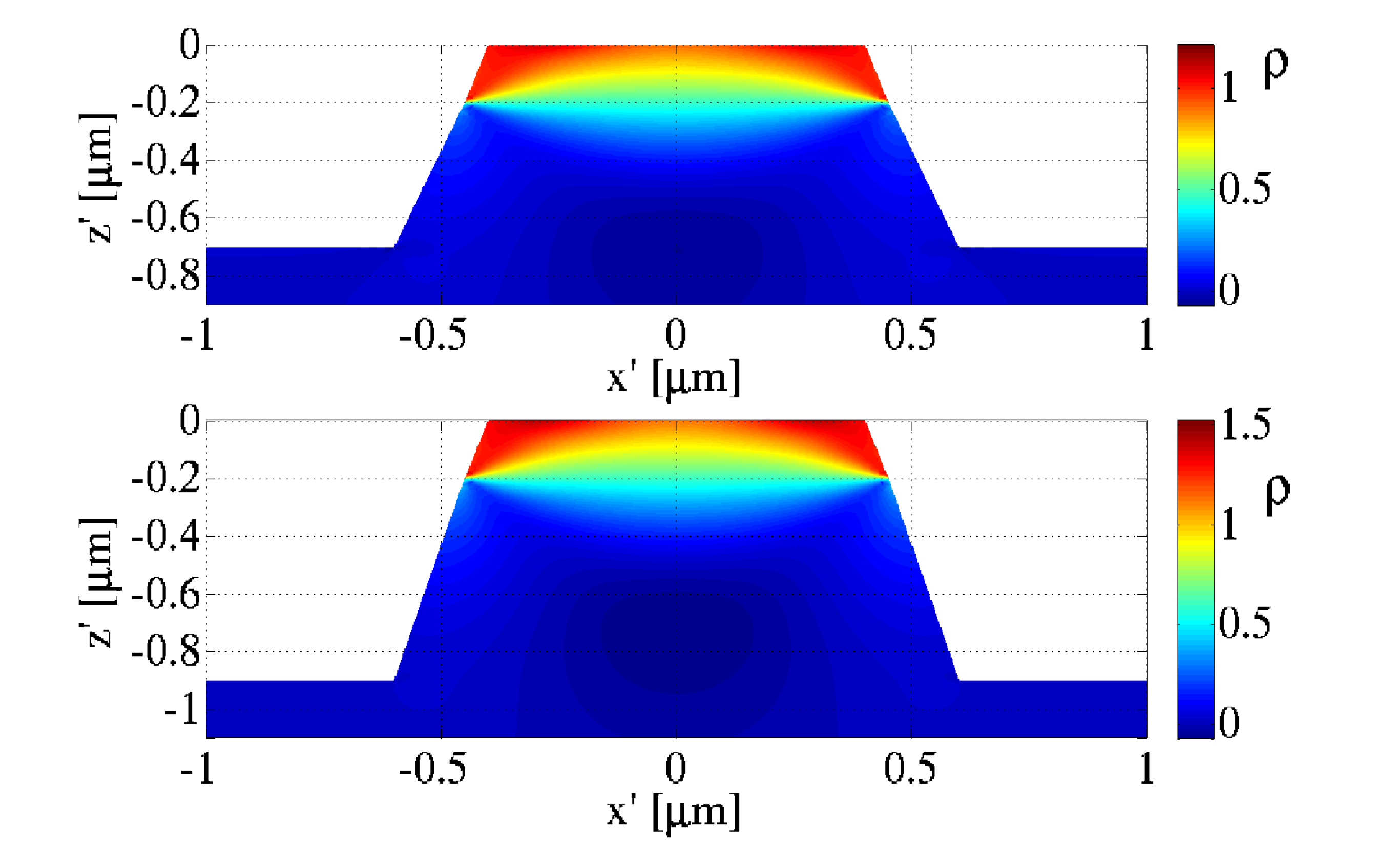}
\caption{(Color online) Finite element calculation of the lattice relaxation, $\rho(x',z')$, on the cross-section perpendicular to the slightly overcut stripes B$_{[1\overline{1}0]}$ (upper panel) and B$_{[110]}$ (lower panel). The cross-section of one stripe and the underlying substrate is plotted. The relaxation $\rho=1$ and $\rho=0$ corresponds to a full relaxation of the lattice and to a lattice under a compressive strain of the unpatterned layer, respectively. Both stripes are close to full relaxation.}
\label{strain_simB}
\end{figure}

The remarkable agreement of the measured and simulated diffraction maps shows that our model of the lattice deformations is quantitatively relevant in determining the local lattice relaxation $\rho(x',z')$ in the stripes shown in Fig.~\ref{strain_simB}, the lattice mismatch between the epilayer and the substrate, $e_0=-0.38\pm 0.03\%$ for set~B, and the stripe geometry, a trapezoidal cross-section of the stripe also shown in Fig.~\ref{strain_simB}. The largest relaxation is observed in the corners of the stripes.

The slopes of the sides in set~B are few degrees larger than angles typically occurring when dry etching is used during the patterning process. Note that the X-ray diffraction reveals only the regions with regular lattice structure whereas the dry etching can leave a thin non-uniform amorphous coating on the stripes which leads to the unexpected non-rectilinear shape of the stripe cross-section resulting from the fitting.

\begin{figure}
\includegraphics[scale=0.3]{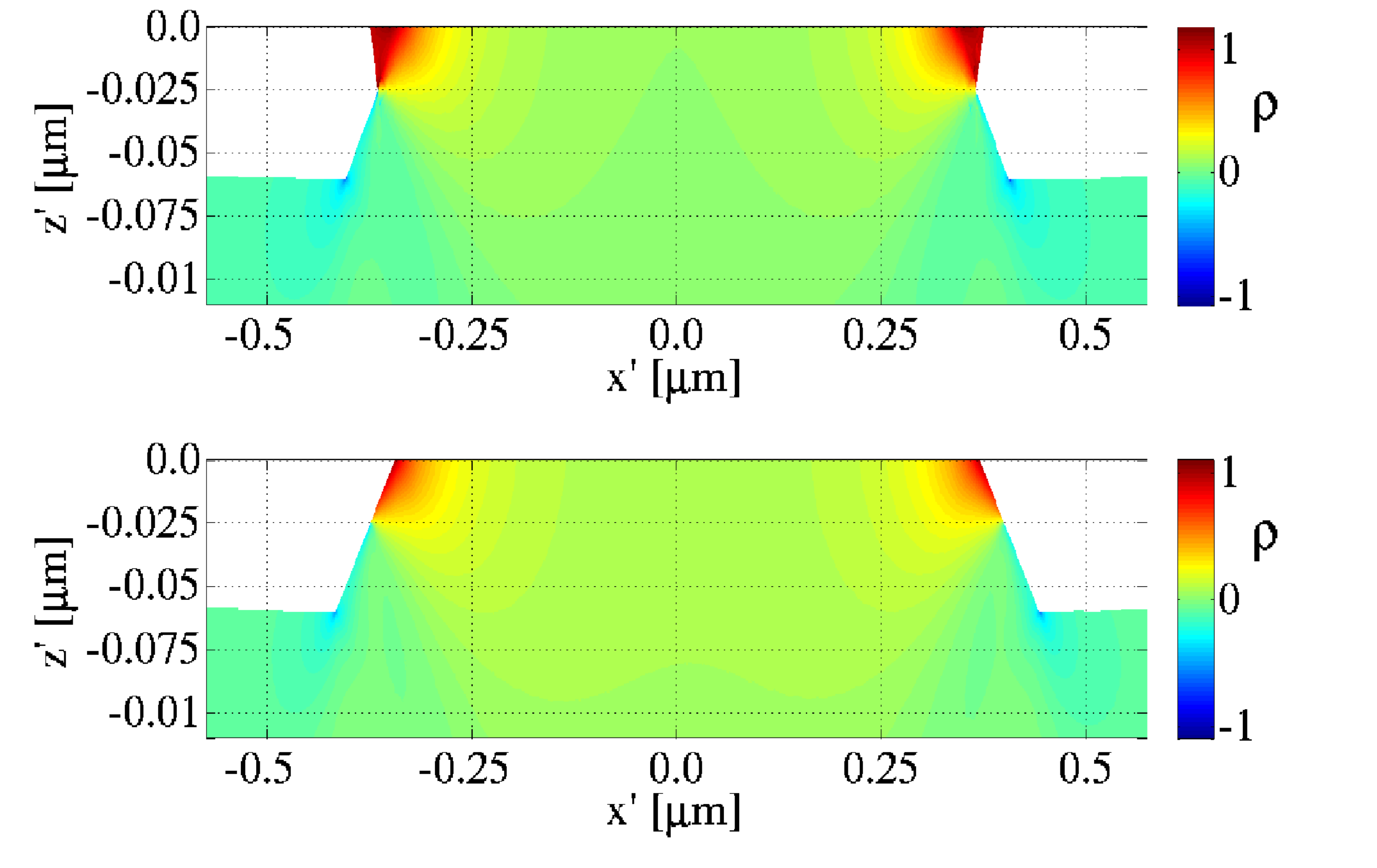}
\caption{(Color online) Finite element calculation of the lattice relaxation, $\rho(x',z')$, on the cross-section perpendicular to the undercut stripes A$_{[110]}$ (upper panel) and overcut stripes A$_{[1\overline{1}0]}$ (lower panel). The cross-section of one stripe and the underlying substrate is plotted. The relaxation $\rho=1$ and $\rho=0$ corresponds to full relaxation of the lattice and to a lattice under a compressive strain of the unpatterned layer, respectively. All stripes show weaker net relaxation than the stripes in set~B.}
\label{strain_simA}
\end{figure}

In the next step, we use our modelling of the lattice relaxation also for stripes of set~A where the X-ray diffraction would be less accurate due to the small volume of the epilayer, however, the relaxation mechanism should be of the same nature as in set~B. Fig.~\ref{strain_simA} shows the spatial dependence of the function $\rho(x',z')$ for two different geometries relevant to samples in set~A. The shape of the stripe cross-section cannot be determined from the SEM image of Fig.~\ref{sem_Nott} with nanometer accuracy. This uncertainty cannot be neglected in the undercut stripes A$_{[110]}$. Therefore, more geometries (slopes of the sides) were simulated and one representative example is given in the upper panel of Fig.~\ref{strain_simA}. On the other hand, the precise shape of the sides does not play such an important role in case of the overcut stripes A$_{[1\overline{1}0]}$ shown  in the lower panel of Fig.~\ref{strain_simA}. In all geometries, the local induced strain is stronger closer to the edges of the stripes.

The comparison of the macroscopic simulations and X-ray diffraction measurements are done on the level of the full spatial distribution of the relaxation $\rho(x',z')$. The magnetic characteristics, considered in this work in the single domain approximation,  are analyzed based on the net lattice relaxation. Here we take advantage of the direct proportionality of the magnetocrystalline anisotropy to the corresponding strain\cite{Daeubler:2007_a,Zemen:2009_a} and calculate the mean anisotropy from the spatial average of $\rho(x',z')$ over the stripe cross-section. We will denote this average quantity by $\hat{\rho}$ in the rest of the paper.

The last step in obtaining the input parameters for the microscopic modelling is writing the net in-plane components of the total strain tensor introduced in Eq.~(\ref{strain_relax_comsol}) in the coordinate system fixed to the main crystallographic axes introduced in Fig.~\ref{sem_Nott}:
\begin{eqnarray}
\label{strain_matrix}
e_{xx} & = & e_0\left(1-\frac{\hat{\rho}}{2}-\frac{\hat{\rho}}{2}\sin 2\omega\right), \\
e_{yy} & = & e_0\left(1-\frac{\hat{\rho}}{2}+\frac{\hat{\rho}}{2}\sin 2\omega\right), \nonumber \\
e_{xy} & = & e_0\frac{\hat{\rho}}{2}\cos 2\omega, \nonumber \;
\end{eqnarray}
where the angle $\omega$ is measured from the $[1\overline{1}0]$ axis and the angle $\omega-45^{\circ}$ describes the rotation of the relaxation direction (the dashed coordinates) with respect to the crystalline coordinate system. Note that the above strain components coincide with those in Eq.~(\ref{strain_relax_comsol}) when $\omega=45^{\circ}$, i.e., the relaxation direction is aligned with the $[100]$ axis. We emphasize that the average relaxation $\hat{\rho}$ depends on $\omega$. We rotate the elasticity matrix describing the cubic crystal when simulating the lattice relaxation along different directions.

The strain components $e_{xx}$, $e_{yy}$, and $e_{xy}$ for the stripes in set~A are obtained from the macroscopic simulations and considering $e_0\approx-0.22\%$.\cite{Zhao:2005_a,Jungwirth:2005_b}
\begin{table} [h]
\begin{tabular}{|c|c|c|} \hline
                       & $e_0 [\%]$     & $\hat{\rho}$      \\ \hline 
A$_{[1\overline{1}0]}$ & $-0.22\pm 0.03$ & $0.184\pm 0.005$ \\ \hline 
A$_{[110]}$            & $-0.22\pm 0.03$ & $0.24\pm 0.05$   \\ \hline
B$_{[1\overline{1}0]}$ & $-0.38\pm 0.03$ & $0.79\pm 0.01$   \\ \hline 
B$_{[010]}$            & $-0.38\pm 0.03$ & $0.99\pm 0.01$   \\ \hline 
\end{tabular}
\caption{The lattice mismatch $e_0$ and the lattice relaxation $\hat{\rho}$ for the patterned samples as entering the microscopic calculations in Sec.~\ref{se_theory}. The value of $e_0$ in set~B is determined from the X-ray diffraction experiment, whereas $e_0$ in set~A is inferred from the partial Mn concentrations using the analysis of Refs.~[\onlinecite{Zhao:2005_a},\onlinecite{Jungwirth:2005_b}].}
\label{tab_strain}
\end{table}
Table~\ref{tab_strain} summarizes the parameters determined in this section. 

\section{Experimental magnetic anisotropies}
\label{se_measuredaniso}
In-plane magnetic anisotropies in thin (Ga,Mn)As films are often analyzed using the lowest order decomposition of the free energy profile into separate terms of distinct symmetry.\cite{Liu:2005_d,Welp:2003_a,Thevenard:2007_a} In this study, we follow this track by adopting the following phenomenological formula:
\begin{equation}
\label{free_en}
F(\psi)=-\frac{K_{c}}{4} \sin^2 2\psi+K_{u} \sin^2\psi - K_{\Omega} \sin^2(\psi-\Omega). 
\end{equation}
The cubic symmetry of the underlying zinc-blende structure is described by the first term with minima along the $[100]$ and $[010]$ directions in case of $K_c>0$. The second term quantified by the coefficient $K_u$ describes the so called ``intrinsic'' uniaxial anisotropy along the crystal diagonals present in the unpatterned (Ga,Mn)As epilayers. The last term quantified by $K_{\Omega}$ describes the uniaxial anisotropy with an extremum at an angle $\Omega$ induced by the relaxation of the lattice mismatch of the doped epilayer and the substrate. The angle $\Omega$ is in general not equal to the angle of the corresponding lattice relaxation $\omega$.\cite{Zemen:2009_a} Both angles are measured from the $[1\overline{1}0]$ axis.

\subsection{Remanent magnetization}
Remanent magnetization along the main crystallographic directions was measured by SQUID for both sets of samples. The obtained values include the magneto-crystalline anisotropies described in the previous paragraph as well as the shape anisotropy which always prefers the magnetization alignment with the longest side of a rectangular prism such as the stripes.\cite{Aharoni:1997_a}

\begin{figure}
\includegraphics[scale=0.4]{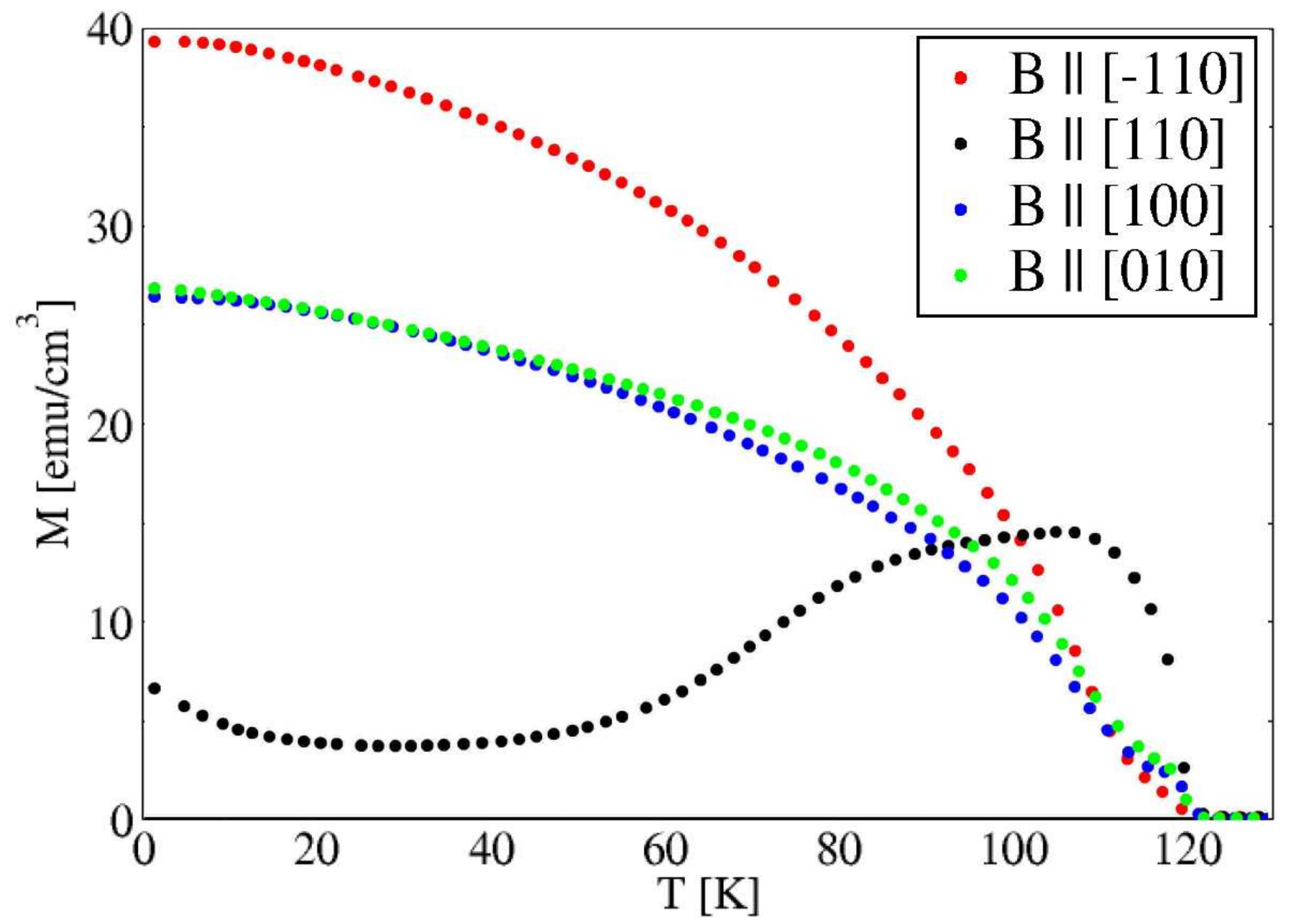}
\caption{(Color online) Remanent magnetization along the main crystallographic directions for sample A$_0$ (25~nm thick unpatterned epilayer).}
\label{rem_Nott_bulk}
\end{figure}

Fig.~\ref{rem_Nott_bulk} shows that in the control sample A$_0$ the intrinsic uniaxial anisotropy dominates over the cubic anisotropy on a large temperature range and the easy axis along the $[1\overline{1}0]$ diagonal. The ratio of the remanent magnetization projections to the $[1\overline{1}0]$ and $[100]$ directions below 60~K reveals that the system is almost purely uniaxial. The behavior of the anisotropy components at $T>60$~K cannot be described within the single domain approximation. However, the anisotropies of unpatterned samples are relevant to our microscopic analysis of measurements in the  microbars only at the lowest temperatures where we extract intrinsic anisotropy coefficients and deduce the material parameters as detailed in Sec.~\ref{se_theory}.

\begin{figure}
\includegraphics[scale=0.4]{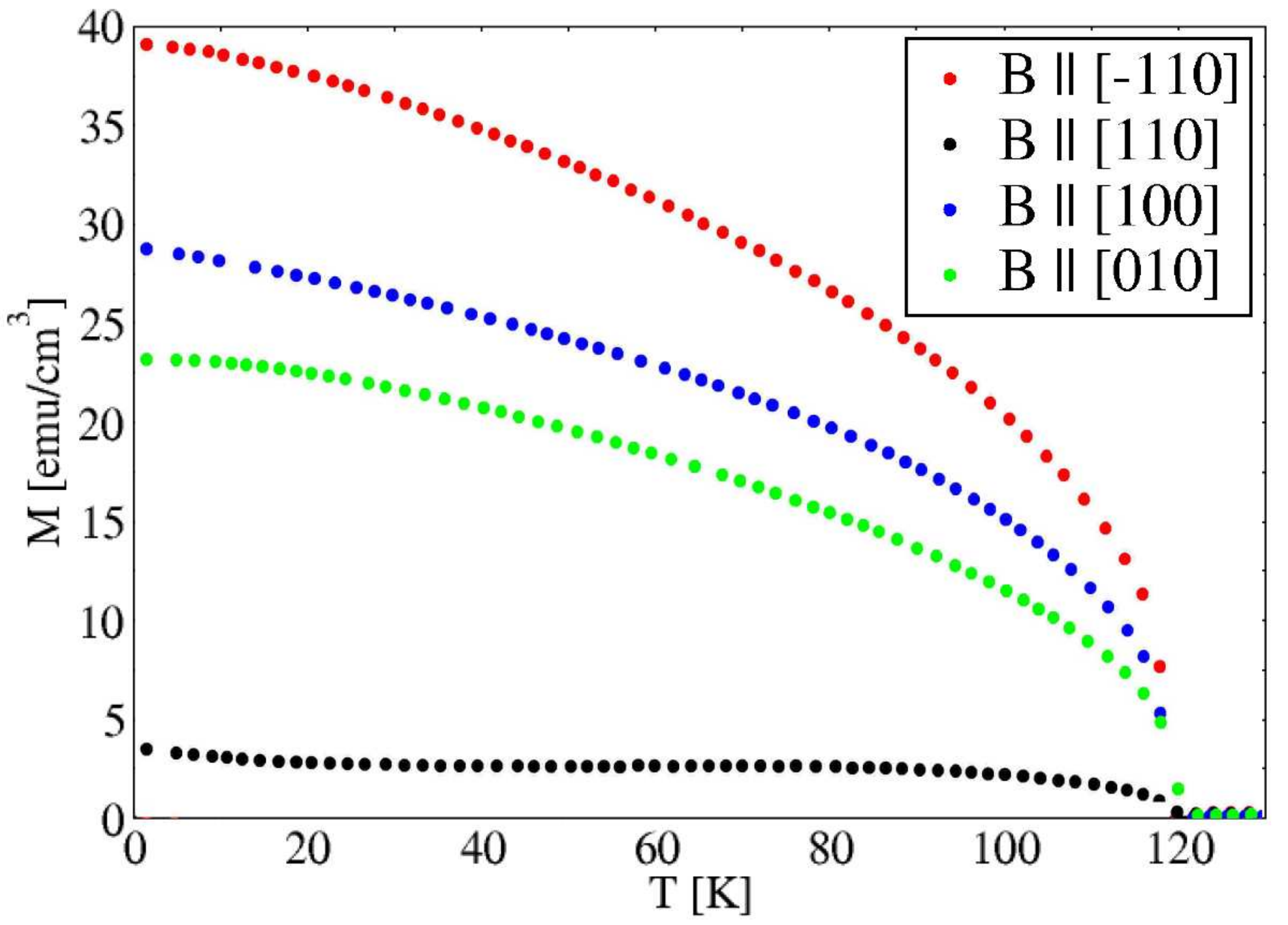}
\caption{(Color online) Remanent magnetization along the main crystallographic directions for sample A$_{[1\overline{1}0]}$ (750~nm wide stripes along the $[1\overline{1}0]$ direction).}
\label{rem_Nott_110}
\end{figure}

Fig.~\ref{rem_Nott_110} shows that the patterning of the sample A$_{[1\overline{1}0]}$ strengthens the uniaxial anisotropy present in the parent wafer. The $[1\overline{1}0]$ diagonal becomes the easiest of the investigated directions at all temperatures and the $[110]$ diagonal becomes the hardest axis at all temperatures below $T_C$. 

\begin{figure}
\includegraphics[scale=0.4]{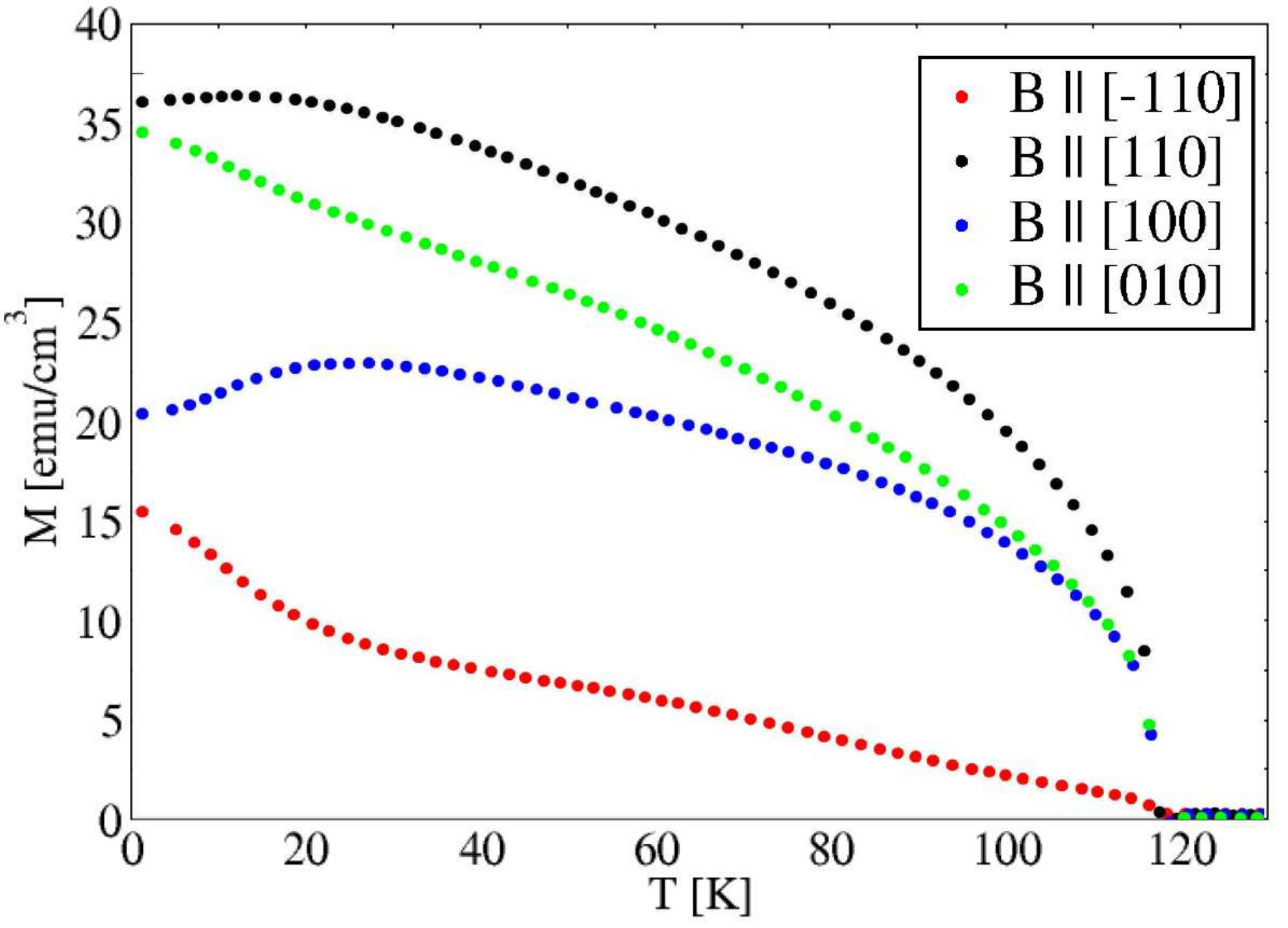}
\caption{(Color online) Remanent magnetization along the main crystallographic directions for sample A$_{[110]}$ (750nm wide stripes along the $[110]$ direction).}
\label{rem_Nott110}
\end{figure}

Fig.~\ref{rem_Nott110} shows that in the sample A$_{[110]}$, the two diagonals switch roles and in analogy with the previous case the easy axis prefers alignment close to the stripe direction, which is the hard axis over most of the temperature range in the parent wafer. This means that  a rotation of the easy axis by as much as 90$^{\circ}$ is achieved by the post-growth patterning.
Note that the difference of the projection of the remanent magnetization to the $[100]$ and $[010]$ directions in the two patterned samples is due to a 5$^{\circ}$ misalignment between the stripes and the crystal diagonals introduced during the fabrication.

\begin{figure}
\includegraphics[scale=0.95]{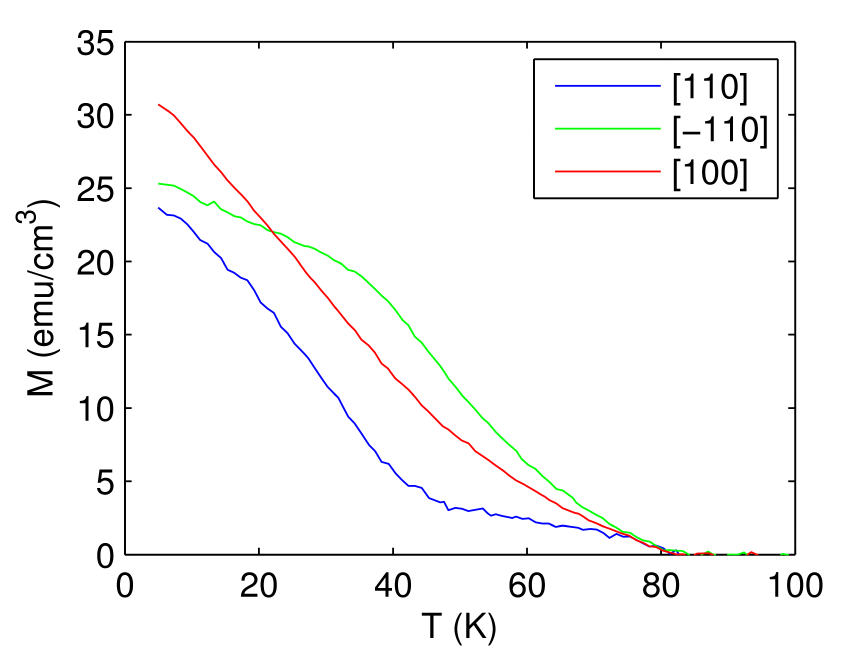}
\caption{(Color online) Remanent magnetization along the main crystallographic directions for sample B$_0$ (200~nm thick unpatterned epilayer).}
\label{rem_Prg_bulk}
\end{figure}

The samples in set~B posses stronger cubic anisotropy. Fig.~\ref{rem_Prg_bulk} shows that in the control sample B$_0$ the intrinsic uniaxial anisotropy dominates over the cubic anisotropy only at temperatures above 20~K and the $[1\overline{1}0]$ diagonal is easier than the $[110]$ diagonal at all temperatures below $T_C$.

\begin{figure}
\includegraphics[scale=0.95]{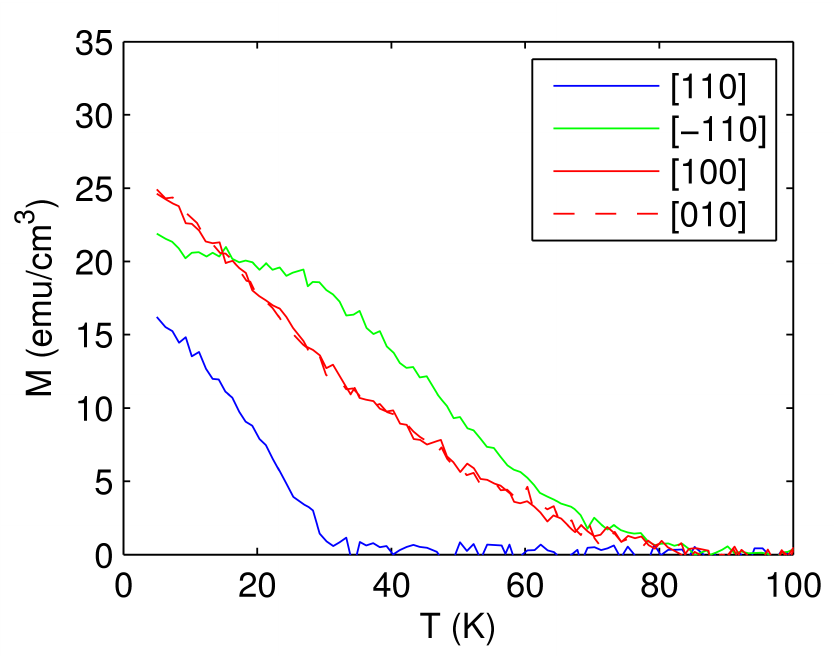}
\caption{(Color online) Remanent magnetization along the main crystallographic directions for sample B$_{[1\overline{1}0]}$ (1~$\mu$m wide stripes along the $[1\overline{1}0]$ direction).}
\label{rem_Prg_110}
\end{figure}

Fig.~\ref{rem_Prg_110} shows a strengthening of the uniaxial anisotropy along the stripe direction in the sample B$_{[1\overline{1}0]}$, although not large enough to overcome the cubic anisotropy at the lowest temperatures. The transition from cubic to uniaxial anisotropy occurs at a lower temperature than in the control sample. The $[110]$ direction is hardened. The main crystal axes $[100]$ and $[010]$ remain equal due to the more accurate alignment of the stripes with the crystal diagonal.

\begin{figure}
\includegraphics[scale=0.95]{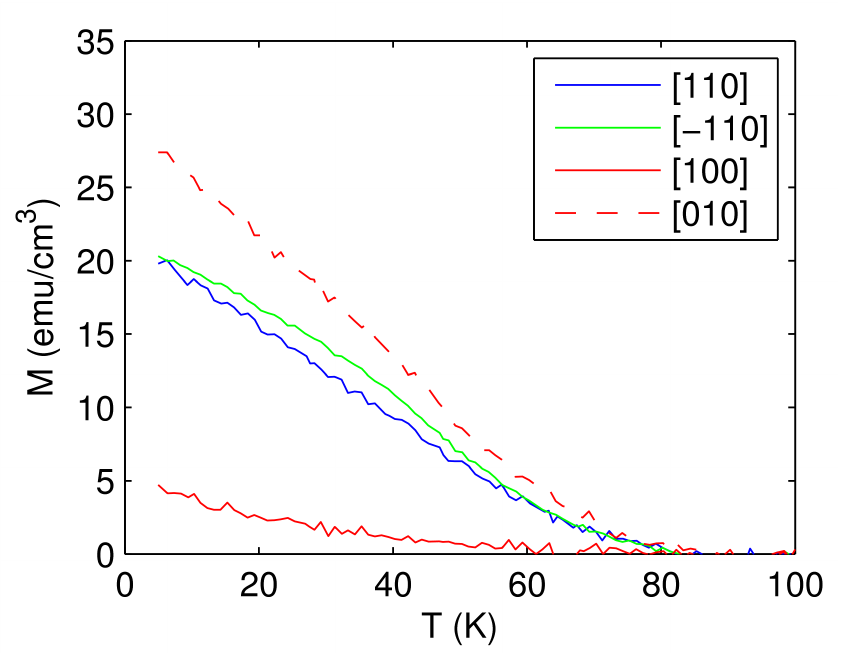}
\caption{(Color online) Remanent magnetization along the main crystallographic directions for sample B$_{[010]}$ (1~$\mu$m wide stripes along the $[010]$ direction).}
\label{rem_Prg010}
\end{figure}

Fig.~\ref{rem_Prg010} shows a differentiation of the $[100]$ and $[010]$ projections in the sample B$_{[010]}$. The uniaxial anisotropy along the stripe direction now dominates at all temperatures. The intrinsic anisotropy differentiating the diagonal directions is less pronounced than in case of B$_0$ as it has to compete also with the induced uniaxial anisotropy.

We can conclude that the universal effect seen in all patterned (Ga,Mn)As/GaAs samples is the preference of the easy axis to align parallel to the  stripe which is the direction in which the growth induced compressive strain cannot relax, i.e., the direction of the relative lattice contraction in (Ga,Mn)As. This is reminiscent of the magnetocrystalline anisotropy of unpatterned (Ga,Mn)As epilayers which typically yields easy-axis oriented also along the direction of contraction, i.e., in-plane for compressively strained (Ga,Mn)As epilayers and out-of-plane for (Ga,Mn)As films grown under tensile strain.\cite{Zemen:2009_a} We point out that the measured magnitudes of magnetic anisotropies in the microbars are an order of magnitude larger than the shape anisotropy contribution for given concentration of magnetic moments and thickness to width ratio. The microfabrication effects in the (Ga,Mn)As stripes are therefore primarily due to the spin-orbit coupling induced magnetocrystalline anisotropy.

\subsection{Anisotropy coefficients}
After investigating the reorientations of the easy axis we focus on the magnitude of the individual anisotropy components. We measure the hysteresis loops using the SQUID magnetometry and fit the results to the following equation: 
\begin{eqnarray}
\label{free_en_fit}
F(\psi)/\mu_0 & = & -\frac{1}{4} M_S H_c \sin^2 2\psi + M_S H_u \sin^2\psi - \\
 & & - M_S H_{\Omega} \sin^2(\psi-\Omega) - M_S H cos(\psi-\phi_H), \nonumber \;
\end{eqnarray}
where $K_i = \mu_0 M_S H_i$ were introduced in Eq.~(\ref{free_en}), $M_S$ is the saturation magnetization, $H$ is the external magnetic field applied at the angle $\phi_H$, and the last term is the Zeeman energy. All angles in Eq.~(\ref{free_en_fit}) are measured from the $[1\overline{1}0]$ axis. In case of a general alignment of the induced uniaxial strain, the angle $\Omega$ of the corresponding uniaxial anisotropy is an independent fitting parameter. However, in case of the main crystallographic axes and their small surrounding we can set $\Omega=\omega$, i.e., the anisotropy term is aligned with the corresponding uniaxial strain.\cite{Zemen:2009_a} An overview of the resulting angles $\Omega$ for the different alignments of stripes in sets A and B is given in Table~\ref{tab_coef}.
\begin{table} [h]
\begin{tabular}{|c|c|c|c|c|} \hline
                       & $K_c$ [kJ/m$^3$] & $K_u$ [kJ/m$^3$] & $K_{\Omega}$ [kJ/m$^3$] & $\Omega$ [deg] \\ \hline
A$_0$                  & 0.412            & 0.404            & 0.0                     &                \\ \hline
A$_{[1\overline{1}0]}$ & 0.412            & 0.404            & 0.83                    & 95             \\ \hline 
A$_{[110]}$            & 0.412            & 0.404            & 1.037                   & 5              \\ \hline
B$_0$                  & 2.213            & 0.381            & 0.0                     &                \\ \hline
B$_{[1\overline{1}0]}$ & 2.213            & 0.381            & 0.935                   & 90             \\ \hline 
B$_{[010]}$            & 2.213            & 0.381            & 0.696                   & 45             \\ \hline 
\end{tabular}
\caption{The anisotropy coefficients obtained by fitting the hysteresis loops at $T=2$~K to Eq.~(\ref{free_en_fit}) and the angular shift of the anisotropy term induced by the lattice relaxation as introduced in Eq.~(\ref{free_en}). Note that the lattice relaxes perpendicular to the stripe direction. The error of the anisotropy coefficients is approximately $10-20\%$, approaching the upper limit in case of the thick inhomogeneous samples in set~B.}
\label{tab_coef}
\end{table}

When determining the anisotropy coefficients in the stripes we use the assumption of linear superposition of the anisotropies present in the unpatterned samples with the anisotropies induced by the patterning and lattice relaxation: the coefficients $K_c$ and $K_u$ are obtained first in the control samples and kept fixed when fitting the stripes fabricated from the same epilayer. The assumption is justified on the qualitative level by the remanent magnetization measurement discussed in the previous subsection which revealed the persistence of the bulk anisotropies in all patterned samples. Its validity has been corroborated also by studies of epilayers subject to post-growth piezo straining\cite{Rushforth:2008_a} and lithographic patterning.\cite{Wunderlich:2007_c} We emphasize that our approach is appropriate only when the lattice relaxation direction is very close to the main crystallographic axes or when the angle $\Omega$ is also treated as a fitting paramater.\cite{Zemen:2009_a}

Another assumption concerns the magnetization reorientation mechanism determining the shape of the hysteresis loops. In case of a dominant uniaxial anisotropy we fit the hysteresis loops obtained for external fields applied along the hard axis. In case of a dominant cubic anisotropy there is no completely hard direction. We nevertheless still consider a single domain model in the fitting.  

Anisotropy coefficients for all six samples at the lowest temperature are summarized in Table~\ref{tab_coef}. Recall that these energies include also the contribution of the shape anisotropy which amounts to $\sim 0.1$~kJ/m$^3$ in samples A$_{[1\overline{1}0]}$ and A$_{[110]}$ and $\sim 0.3$~kJ/m$^3$ in the samples B$_{[1\overline{1}0]}$ and B$_{[010]}$ with the higher thickness to width ratio. 
Note that the smaller coefficient $K_{45}$ leads to the formation of a strongly uniaxial system  as shown in Figs.~\ref{rem_Prg010}, whereas the larger coefficient $K_{90}$ cannot overcome  the cubic anisotropy component, at least at low temperatures as shown in Fig.~\ref{rem_Prg_110}. It is because in case of sample B$_{[010]}$, the induced anisotropy is added along the $[010]$ axis which was already the easy (together with $[100]$) direction in the unpatterned epilayer.

\begin{figure}
\includegraphics[scale=0.3]{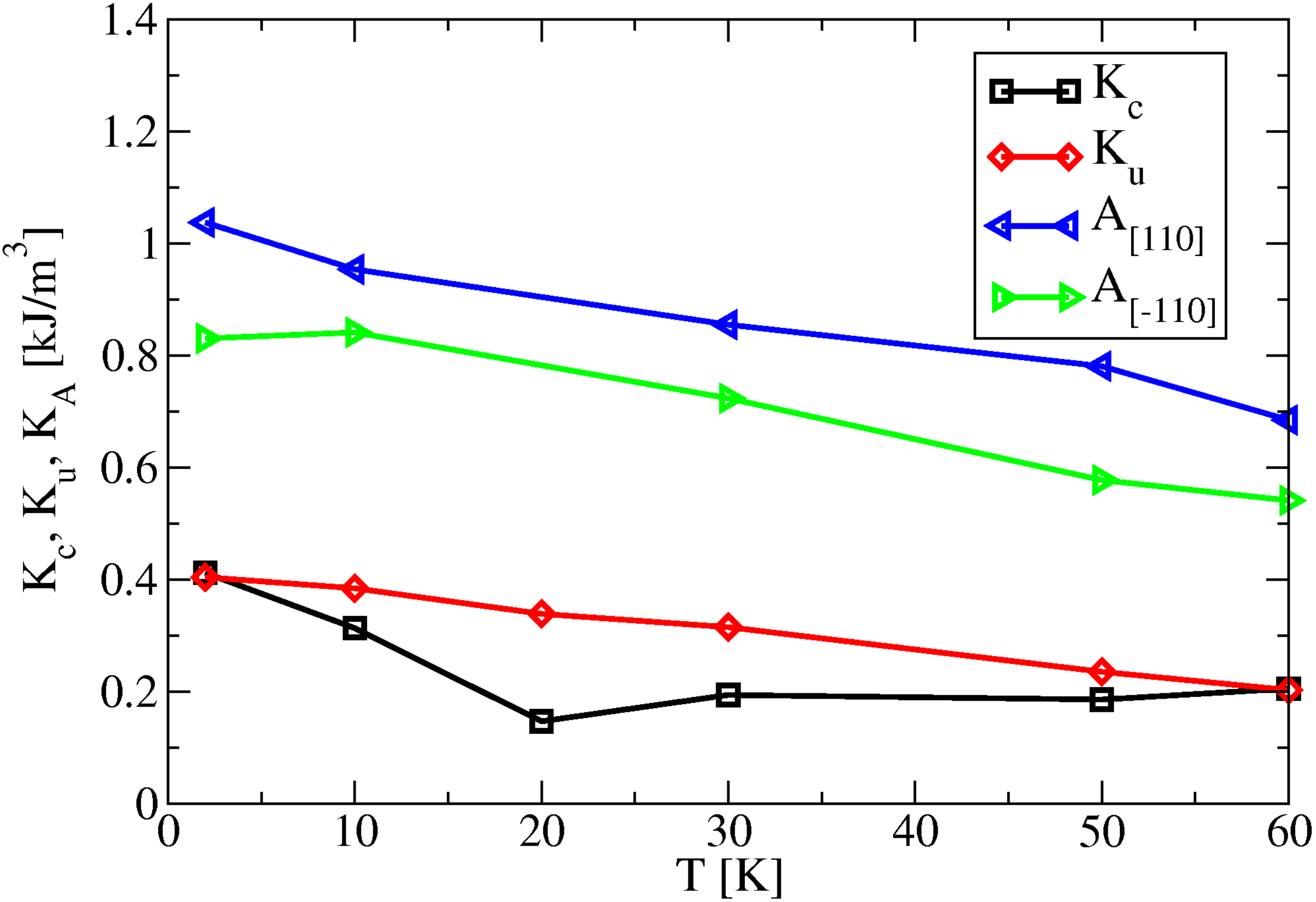}
\caption{(Color online) Anisotropy coefficients as functions of temperature obtained by fitting the hysteresis loops to Eq.(\ref{free_en}) for the three samples of set~A. The uniaxial coefficients $K_{\Omega}$ (denoted by $K_A$ for set~A) due to the growth strain relaxation in the patterned samples dominate the total anisotropy.}
\label{coef_exp_Nott}
\end{figure}

For the thinner and more homogeneous epilayers in set~A we were able to extract the temperature dependence of the anisotropy coefficients from the hysteresis loops up to $T=60$~K as shown in Fig.~\ref{coef_exp_Nott}. The uniaxial coefficients due to lattice relaxation dominate the anisotropy at all temperatures. At low temperatures the relative size of the induced anisotropies corresponds well to the simulated relaxations $\hat{\rho}$: Sample A$_{[1\overline{1}0]}$ with overcut sides (weaker relaxation) shows smaller anisotropy than sample A$_{[110]}$ with undercut sides (stronger relaxation).
The cubic anisotropy remains positive for all studied temperatures $T<60$~K which is in good agreement with the remanent magnetization data shown in Fig.~\ref{rem_Nott_bulk}. We do not discuss measurements above 60~K for which, as mentioned above, the single domain model is not applicable. 

\section{Comparison with theory}
\label{se_theory}

In this section we build on macroscopic calculations of the lattice relaxation presented in Sec.~\ref{se_strain}, perform the microscopic calculations of the magnetic anisotropy energy, and analyze its correspondence with the experimental results on the level of anisotropy coefficients. We extract the coefficients by fitting the calculated total energies to Eq.~(\ref{free_en}) for different magnetization directions.

The comparison of the experimental and theoretical results involves a number of material parameters. The most important inputs of the microscopic calculations are the concentration of the ferromagnetically ordered Mn local moments ($x$) and the hole density ($p$). Unfortunately, these two parameters cannot be accurately controlled during the growth or determined post growth.\cite{Jungwirth:2005_a} The measured saturation magnetization, the conductivity, and the Curie temperature of the control samples provide only estimates of these input parameters with limited accuracy. 

Another independent input parameter of the microscopic simulations is the ``intrinsic'' shear strain which has been used successfully to model\cite{Sawicki:2004_a,Zemen:2009_a} the in-plane uniaxial anisotropy in the unpatterned samples. We recall that such modelling for small strains (the typical values\cite{Zemen:2009_a} are $e_{in}\sim 10^{-4}$) complies well with the assumption that the ``intrinsic'' uniaxial anisotropy superposes linearly with anisotropy components induced by the lattice relaxation, as mentioned in the previous section. The intrinsic shear strain is added to the off-diagonal element of the total strain tensor written in Eq.~(\ref{strain_matrix}) giving: $e_{xy} = e_{in} + e_0\frac{\hat{\rho}}{2}\cos 2\omega$.

\begin{figure}
\includegraphics[scale=0.3]{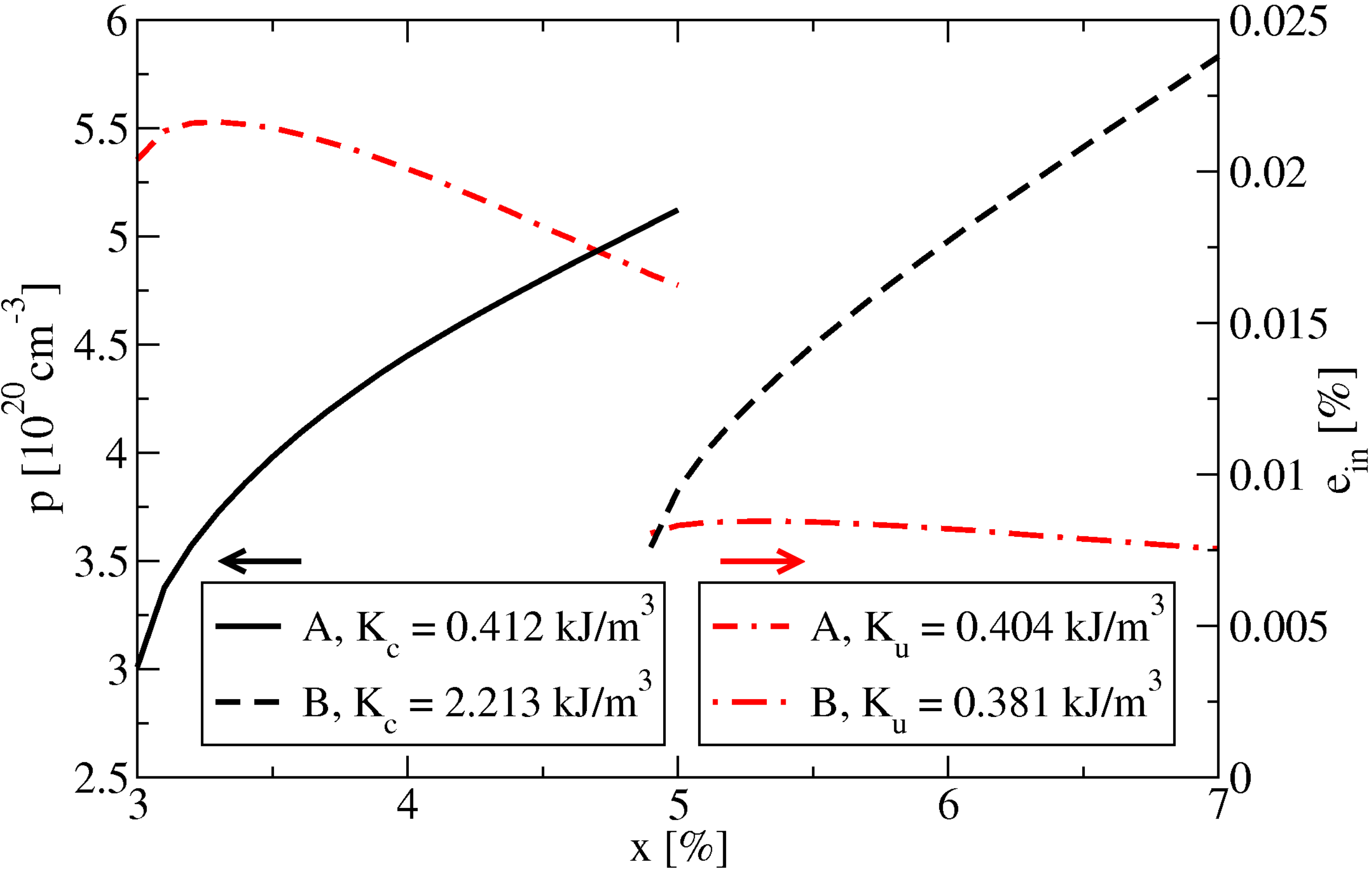}
\caption{(Color online) Correspondence of the hole density, $p$, and the intrinsic shear strain, $e_{in}$, to the effective Mn concentration, $x$, based on the agreement of the calculated $K_c$ and $K_u$ with the measured values. Samples A and B at zero temperature.}
\label{set_xpe}
\end{figure}

Fig.~\ref{set_xpe} shows the combinations of $x$, $p$, and $e_{in}$ for which the calculated intrinsic $K_u$ and $K_c$ of the control samples A$_0$ and B$_0$ agree  with the measured values at zero temperature. By this we limit the intervals of $x$, $p$, and $e_{in}$ values considered in the modeling of the temperature dependent anisotropy coefficients in all measured samples. Note, that we have also imposed an upper bound to $x$ given by the nominal Mn doping in the particular material and a bound to $p$ ensuring a maximum of one hole per Mn ion and in-plane easy axis (axes). This method allows for predicting the induced anisotropy coefficients in the microbars without any adjustable parameters in the microscopic model.

\subsection{Low temperatures}
\begin{figure}
\includegraphics[scale=0.3]{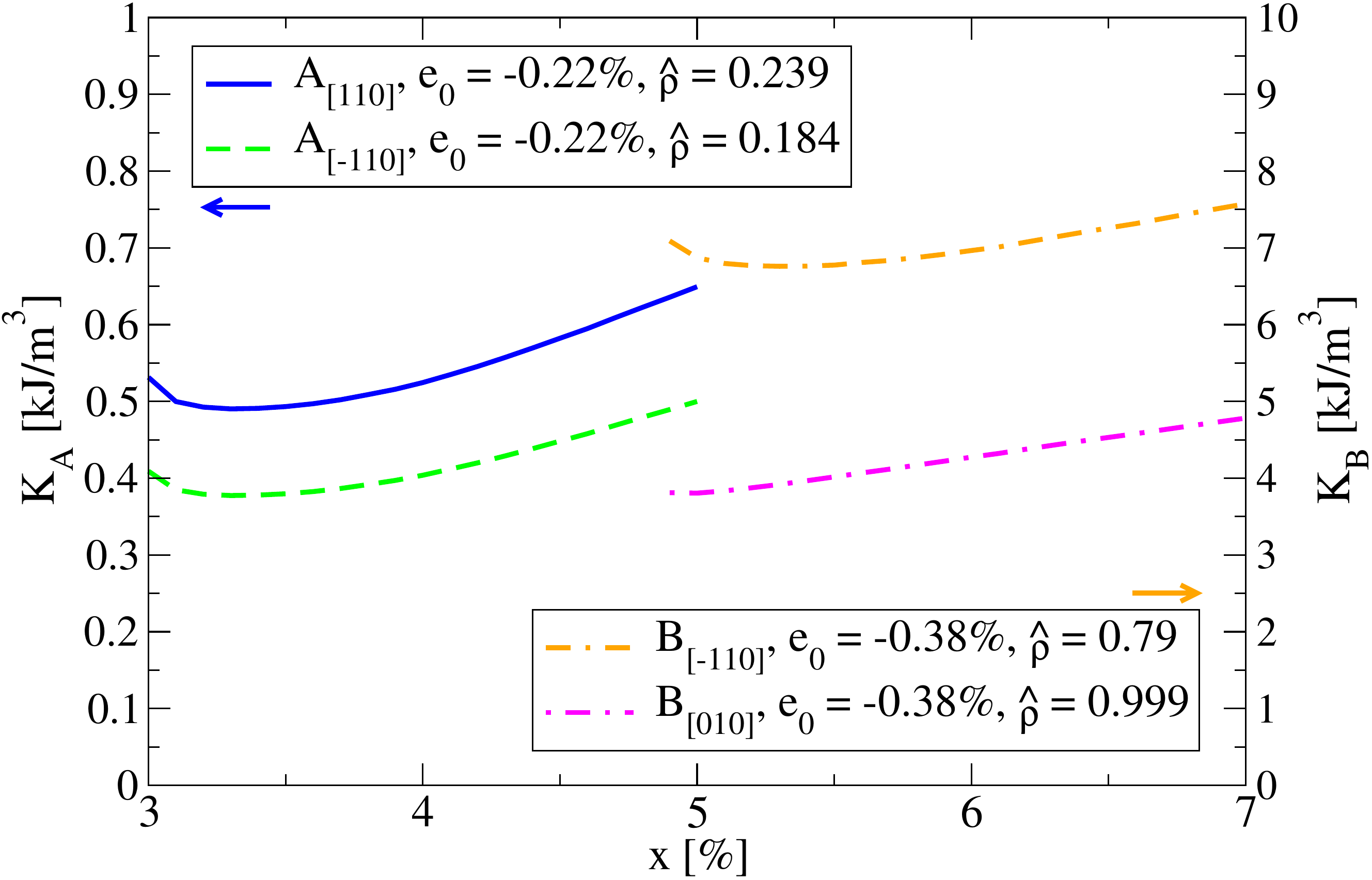}
\caption{(Color online) Calculated anisotropy coefficients due to the lattice relaxation in the patterned samples A and B at zero temperature for fixed combinations of $x$ and $p$ shown in Fig.~\ref{set_xpe}.}
\label{coef_relax_T0}
\end{figure}

Using parameter combinations shown in Fig.~\ref{set_xpe} we calculate the induced uniaxial anisotropies in the microbars at zero temperature. The left and right vertical axis of Fig.~\ref{coef_relax_T0} shows the extracted anisotropy coefficients for stripes in sets~A and~B, respectively. The combinations of $x$, $p$, and $e_{in}$ are indexed only by $x$ for simplicity. The plotted values can be compared to the measured coefficients summarized in Table~\ref{tab_coef}. The relations $K_{95}<K_{5}$ and $K_{90}>K_{45}$ hold both in theory and in experiment. We observe a semi-quantitative agreement in samples A$_{[1\overline{1}0]}$ and A$_{[110]}$ where the measured values are roughly a factor of 2 larger than the calculated ones. The ratio of the calculated coefficients $K_{\Omega}$ for samples A$_{[1\overline{1}0]}$ and A$_{[110]}$, $K_{95}/K_{5}$, is in excellent agreement with experiment (the difference is only $4\%$). These agreements justify the interpretation of the  measured effects in the microbars based on the strain-relaxation controlled magnetocrystalline anisotropy. Note that they also support the  assumption of the linear superposition of  individual uniaxial anisotropies terms used in our analysis.

Fig.~\ref{coef_relax_T0} shows also extracted anisotropy coefficients for samples B$_{[1\overline{1}0]}$ and B$_{[010]}$. In this case, the calculated ratio of coefficients extracted for the two stripe alignments, $K_{90}/K_{45}$, is approximately $20\%$ larger then the corresponding experimental ratio, i.e., still in a very good agreement. We note, however, that the absolute values of the measured coefficients are about a factor of 10 lower than the calculated ones. A possible source of this discrepancy is the large value of the experimentally inferred $K_c$ due to inaccurate subtraction of the paramagnetic and diamagnetic backgrounds from the measured hysteresis curves. In general, we also expect that the theoretical modelling is less reliable in the thicker, as-grown samples B due to stronger disorder and inhomogeneities in the material. 

As a consequence of the almost complete relaxation of the lattice mismatch in the thicker samples the calculated anisotropy coefficients are larger than the cubic coefficient at all studied temperatures which is not in agreement with the measured coefficients in set~B at low temperature (see Table~\ref{tab_coef}).

\subsection{Temperature dependence}
We now select six representative combinations of $x$, $p$, and $e_{in}$ from the relevant interval shown in Fig.~\ref{set_xpe}, calculate the temperature dependence of all anisotropy coefficients for each set of parameters, and discuss the comparison with the measured anisotropies. We recall that in our mean-field modeling at finite temperatures the calculated $T_C$ is uniquely determined by $x$ and $p$. Note that for the entire interval of relevant $x$ and $p$ determined from the low-temperature analysis in the previous section, we obtain Curie temperatures which are in agreement with the experimental values in materials A and B within a factor of 2. This provides an additional support for the overall consistency of our microscopic theoretical analysis of the measured data. 

\begin{figure}
\includegraphics[scale=0.3]{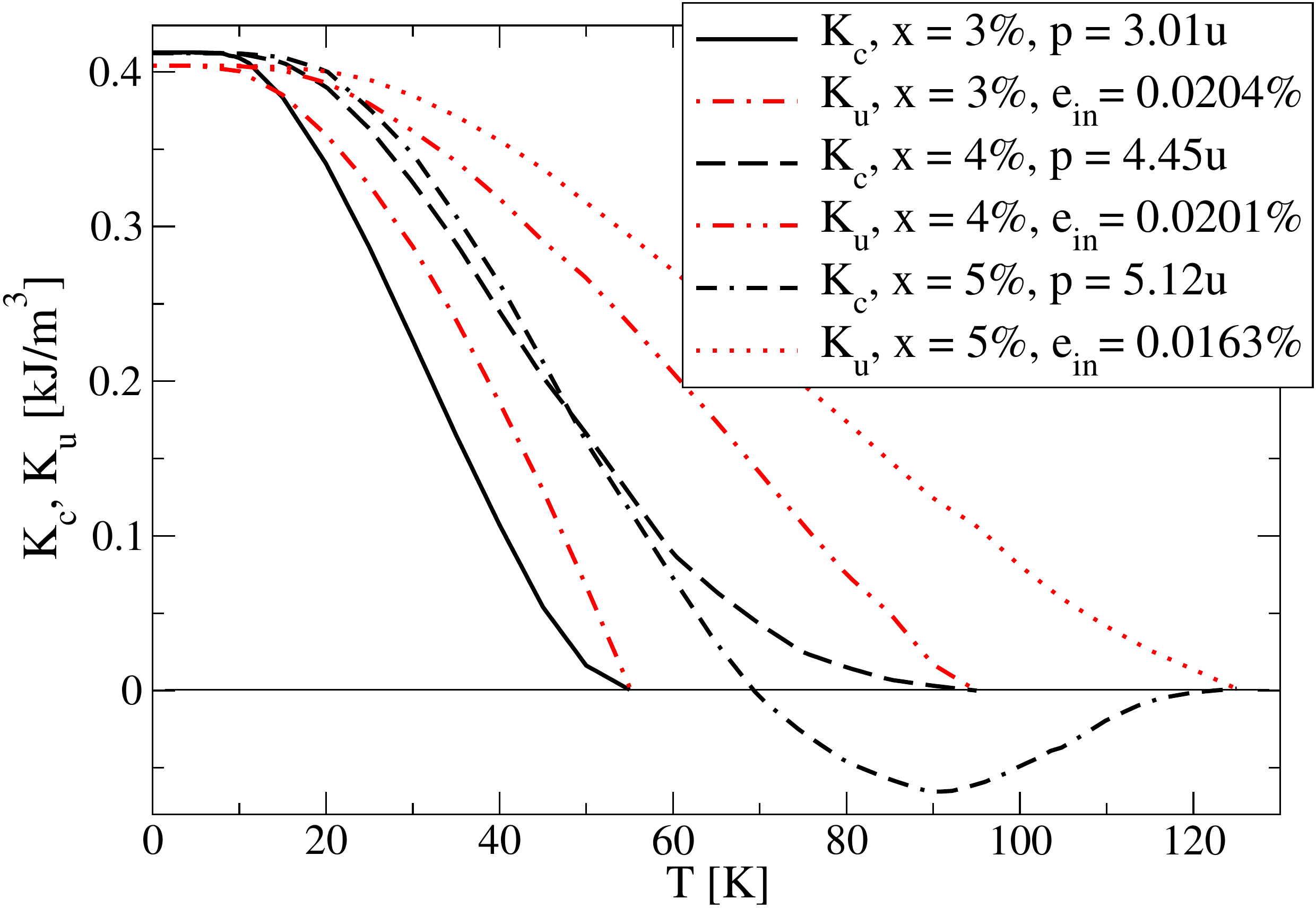}
\caption{(Color online) Calculated cubic a uniaxial intrinsic anisotropy coefficients present in all samples A as functions of temperature for fixed combinations of $x$, $p$, and $e_{in}$ shown in Fig.~\ref{set_xpe}.}
\label{coef_bulk_Tdep_Nott}
\end{figure}

Fig.~\ref{coef_bulk_Tdep_Nott} shows the calculated intrinsic anisotropy coefficients $K_c$ and $K_u$ of samples in set~A for three fixed parameter combinations. At zero temperature the values coincide with data in Fig.~\ref{set_xpe}. The cubic anisotropy component is stronger than the intrinsic uniaxial component at lowest temperatures but it quickly becomes weaker as temperature is increased for all parameter combinations. This temperature dependence is in agreement with the experimental anisotropies measured below 60~K, as shown in Fig.~\ref{coef_exp_Nott}. The comparison cannot be extended to higher temperatures because, as explained above, the experimental behavior at these temperatures is not captured by the single domain model.

\begin{figure}
\includegraphics[scale=0.3]{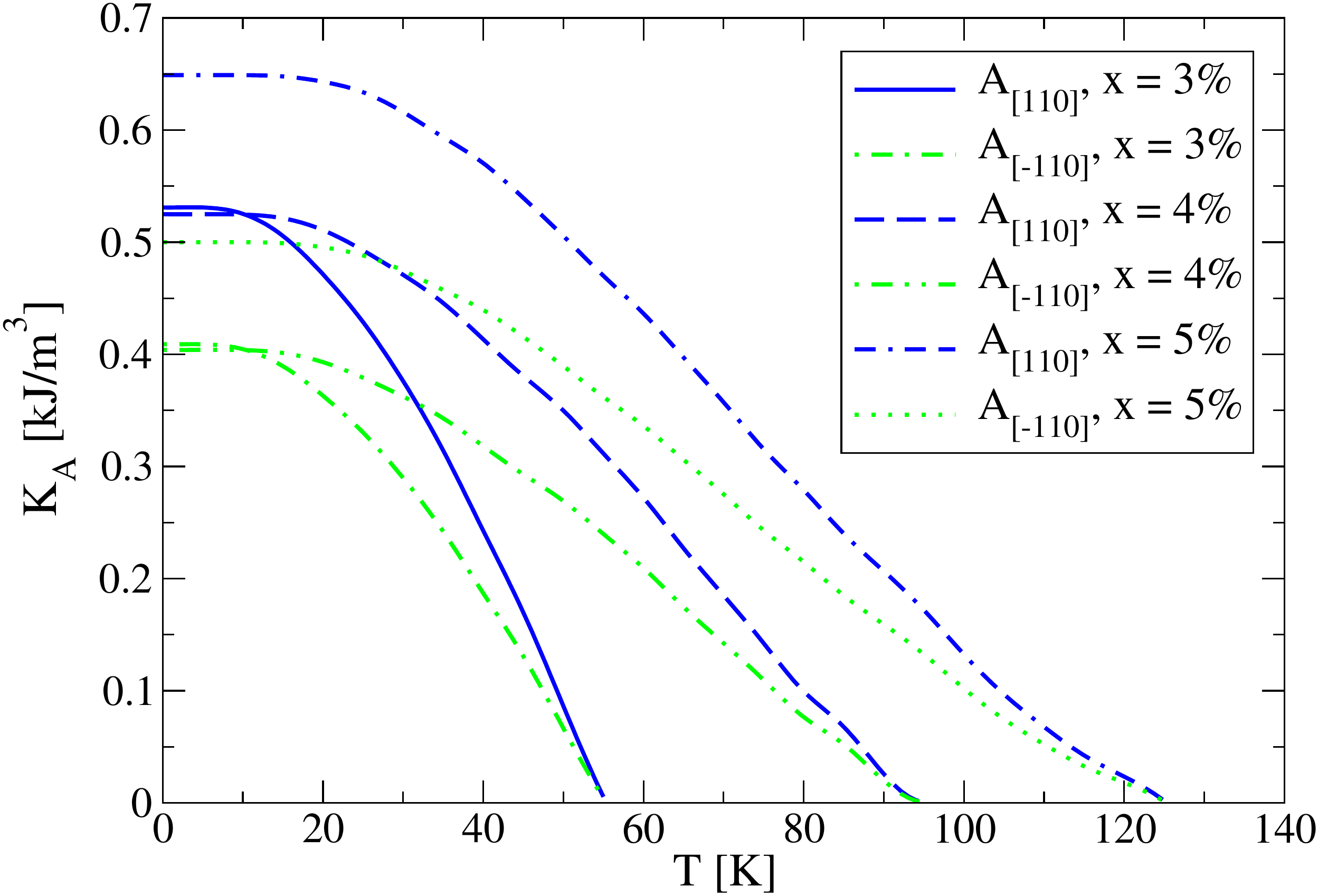}
\caption{(Color online) Calculated anisotropy coefficients due to the lattice relaxation in the patterned samples A as functions of temperature for fixed combinations of $x$ and $p$ shown in Fig.~\ref{set_xpe} and for the induced strain given in Fig.~\ref{coef_relax_T0}.}
\label{coef_relax_Tdep_Nott}
\end{figure}

Fig.~\ref{coef_relax_Tdep_Nott} shows the calculated anisotropy coefficients $K_{\Omega}$ of samples A$_{[1\overline{1}0]}$ and A$_{[110]}$ again for the three fixed parameter combinations. The calculated anisotropy components decrease monotonously with increasing temperature in agreement with the measured dependencies presented in Fig.~\ref{coef_exp_Nott}. The comparison provides additional support for the interpretation of the experimental data, suggested already by the analysis at low-temperature, which is based on the strain relaxation induced magnetocrystalline ansisotropy effects. 

\begin{figure}
\includegraphics[scale=0.3]{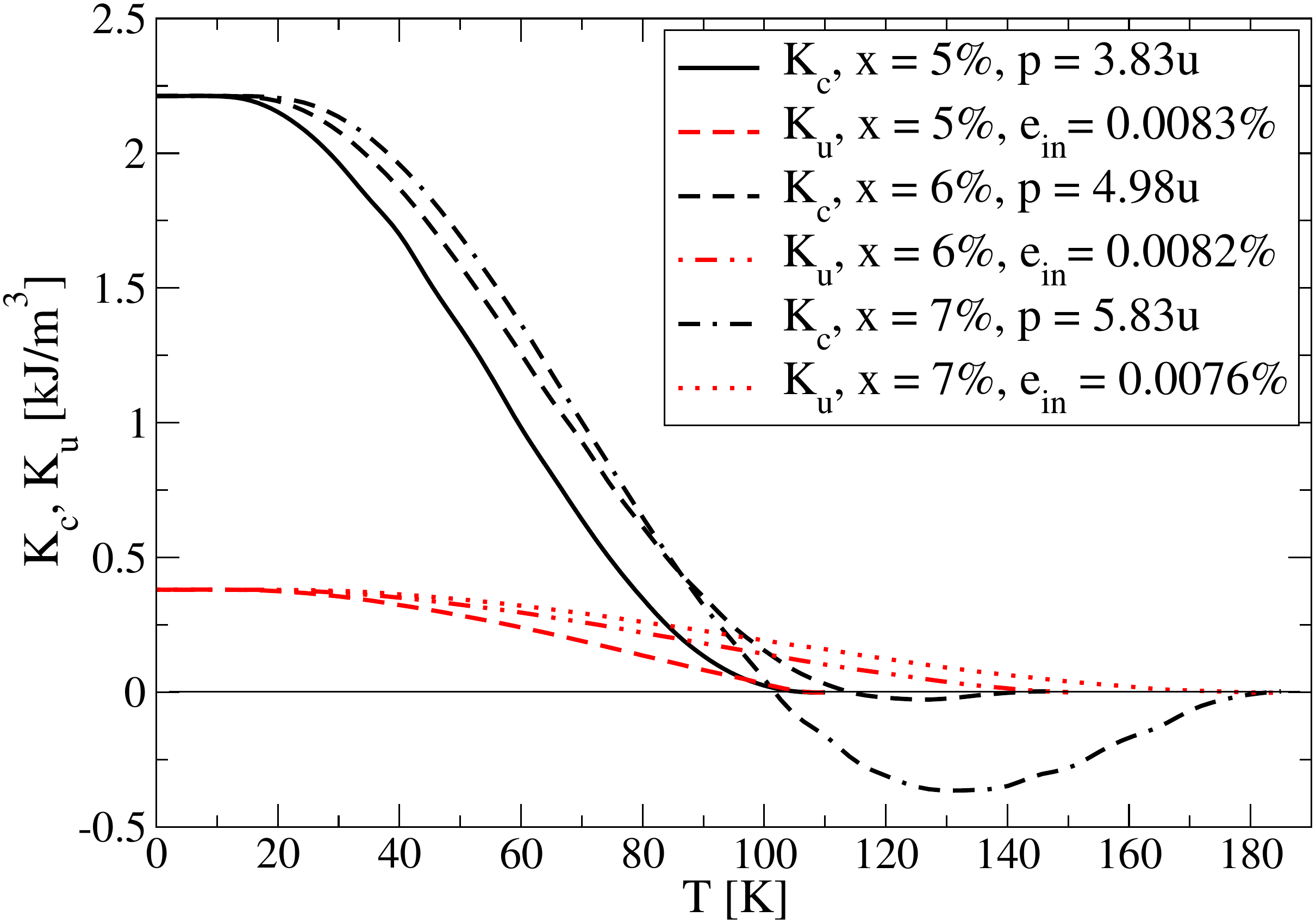}
\caption{(Color online) Calculated cubic a uniaxial intrinsic anisotropy coefficients present in all samples B as functions of temperature for fixed combinations of $x$, $p$, and $e_{in}$ shown in Fig.~\ref{set_xpe}.}
\label{coef_bulk_Tdep_Prg}
\end{figure}

Fig.~\ref{coef_bulk_Tdep_Prg} shows the calculated intrinsic anisotropy coefficients $K_c$ and $K_u$ of samples in set~B again for three fixed parameter combinations. At zero temperature the values coincide with data in Fig.~\ref{set_xpe}. The calculated cubic anisotropy dominates over the uniaxial anisotropy at low temperatures in agreement with the experiment. The cross-over in the theory curves to the dominant uniaxial anisotropy occurs at higher temperatures than $T_C/3$ observed in experiment (see Fig.~\ref{rem_Prg_bulk}); at the upper part of the relevant interval of Mn concentrations the theoretical crossover occurs at $T_C/2$. We again attribute this quantitative discrepancy to inhomogeneities and stronger disorder in the thick as-grown material B.

\begin{figure}
\includegraphics[scale=0.3]{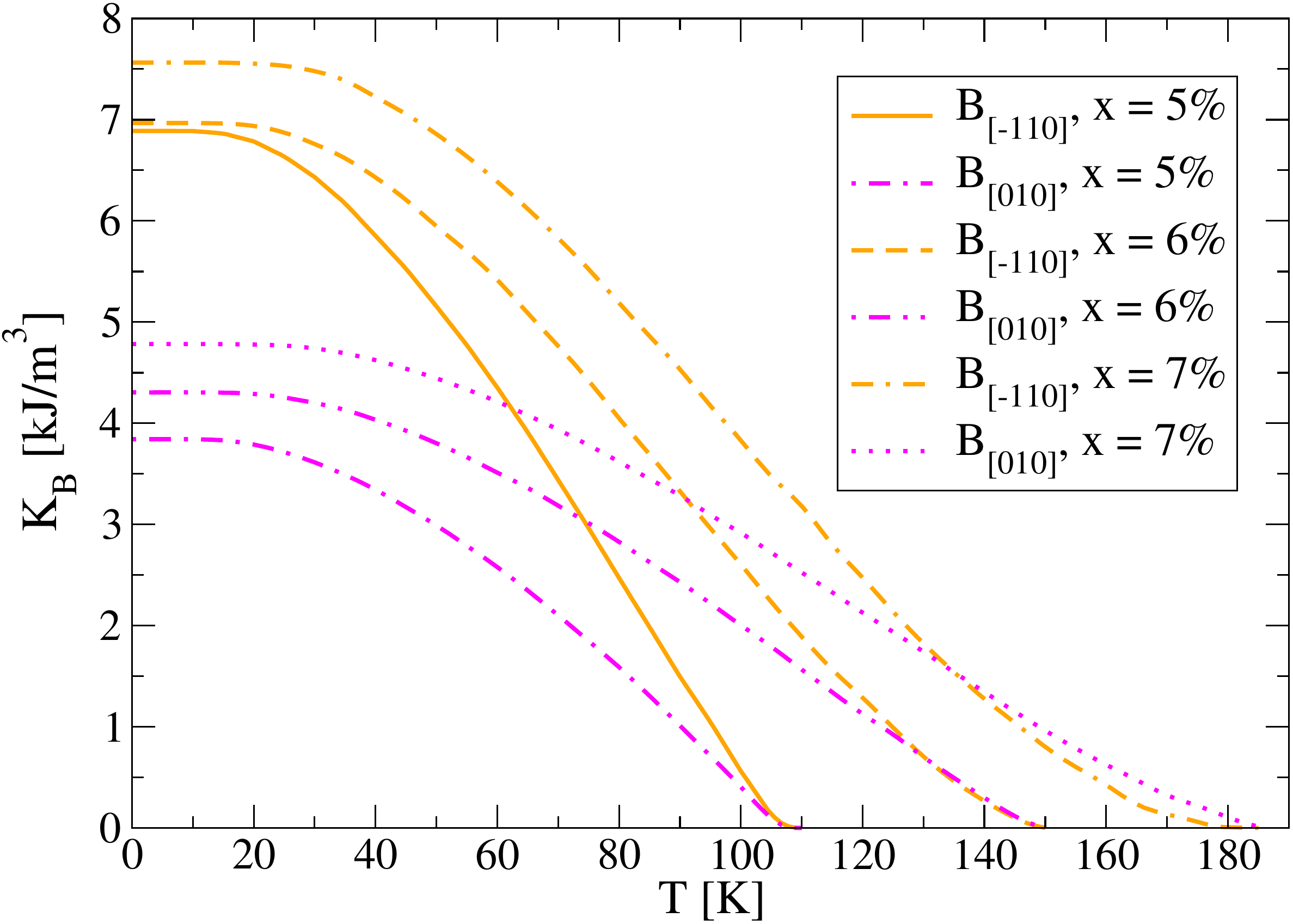}
\caption{(Color online) Calculated anisotropy coefficients due to the lattice relaxation in the patterned samples B as functions of temperature for fixed combinations of $x$ and $p$ shown in Fig.~\ref{set_xpe} and for the induced strain given in Fig.~\ref{coef_relax_T0}.}
\label{coef_relax_Tdep_Prg}
\end{figure}

Fig.~\ref{coef_relax_Tdep_Prg} shows the anisotropy coefficients $K_{\Omega}$ of samples B$_{[1\overline{1}0]}$ and B$_{[010]}$ for the same fixed parameter combinations as in Fig.~\ref{coef_bulk_Tdep_Prg}. Again, we observe very similar dependence of the uniaxial anisotropy coefficients on temperature as in experiment. The monotonous decrease of the coefficients with growing temperature is in agreement with the measured remanent magnetization data in Figs.~\ref{rem_Prg010} and~\ref{rem_Prg_110}. Both induced anisotropy coefficients are predicted to be larger than the cubic coefficient at all studied temperatures. This complies with the measured remanence data of sample B$_{[010]}$. Sample B$_{[1\overline{1}0]}$ shows agreement above 20~K. Its behavior at temperatures below 20~K, is not captured by the theory data as we have already discussed in the previous subsection.

\section{Summary}
\label{se_conclusion}
We have performed a detailed experimental and theoretical analysis of magnetic anisotropies induced in lithographically patterned (Ga,Mn)As/GaAs microbar arrays. Structural properties of the microbars have been studied by X-ray spectroscopy showing strong strain relaxation transverse to the bar axis. The  relaxation induced lattice distortion in stripes with thickness to width ratio as small as $\sim 0.1$ induces additional uniaxial magnetic anisotropy components which dominate the magnetic anisotropy of the unpatterned (Ga,Mn)As epilayer, as revealed by SQUID magnetization measurements. The easy axis can be rotated by the micropatterning by 90$^{\circ}$ at all temperatures below the Curie temperature.

We have carried out systematic macroscopic and microscopic modeling of the structural and magnetic characteristics of the microbars and analyzed in detail the experimental results. The agreement of the measured and simulated X-ray diffraction maps shows that the applied elastic theory model is quantitatively reliable in predicting the local lattice relaxation in patterned epilayers with the growth induced strain.
The overall good agreement of the microscopically calculated and measured magnetic anisotropies conclusively demonstrate that the patterning induced anisotropies are of the magnetocrystalline, spin-orbit coupling origin.

\acknowledgements
We acknowledge fruitful discussions with  A. W. Rushforth and K. V\'{y}born\'{y}. The work was funded through Pr\ae mium Academi\ae{} and contracts number AV0Z10100521, LC510, KAN400100652, FON/06/E002 of \hbox{GA \v CR}, of the Czech republic, and by the NAMASTE (FP7 grant No.~214499) and SemiSpinNet projects (FP7 grant No.~215368).


\begin{thebibliography}{26}
\expandafter\ifx\csname natexlab\endcsname\relax\def\natexlab#1{#1}\fi
\expandafter\ifx\csname bibnamefont\endcsname\relax
  \def\bibnamefont#1{#1}\fi
\expandafter\ifx\csname bibfnamefont\endcsname\relax
  \def\bibfnamefont#1{#1}\fi
\expandafter\ifx\csname citenamefont\endcsname\relax
  \def\citenamefont#1{#1}\fi
\expandafter\ifx\csname url\endcsname\relax
  \def\url#1{\texttt{#1}}\fi
\expandafter\ifx\csname urlprefix\endcsname\relax\def\urlprefix{URL }\fi
\providecommand{\bibinfo}[2]{#2}
\providecommand{\eprint}[2][]{\url{#2}}

\bibitem[{\citenamefont{Edmonds et~al.}(2004)\citenamefont{Edmonds,
  Boguslawski, Wang, Campion, Farley, Gallagher, Foxon, Sawicki, Dietl,
  Nardelli et~al.}}]{Edmonds:2004_a}
\bibinfo{author}{\bibfnamefont{K.~W.} \bibnamefont{Edmonds}},
  \bibinfo{author}{\bibfnamefont{P.}~\bibnamefont{Boguslawski}},
  \bibinfo{author}{\bibfnamefont{K.~Y.} \bibnamefont{Wang}},
  \bibinfo{author}{\bibfnamefont{R.~P.} \bibnamefont{Campion}},
  \bibinfo{author}{\bibfnamefont{N.~R.~S.} \bibnamefont{Farley}},
  \bibinfo{author}{\bibfnamefont{B.~L.} \bibnamefont{Gallagher}},
  \bibinfo{author}{\bibfnamefont{C.~T.} \bibnamefont{Foxon}},
  \bibinfo{author}{\bibfnamefont{M.}~\bibnamefont{Sawicki}},
  \bibinfo{author}{\bibfnamefont{T.}~\bibnamefont{Dietl}},
  \bibinfo{author}{\bibfnamefont{M.~B.} \bibnamefont{Nardelli}},
  \bibnamefont{et~al.}, \bibinfo{journal}{Phys. Rev. Lett.}
  \textbf{\bibinfo{volume}{92}}, \bibinfo{pages}{037201}
  (\bibinfo{year}{2004}).

\bibitem[{\citenamefont{Potashnik et~al.}(2001)\citenamefont{Potashnik, Ku,
  Chun, Berry, Samarth, and Schiffer}}]{Potashnik:2001_a}
\bibinfo{author}{\bibfnamefont{S.~J.} \bibnamefont{Potashnik}},
  \bibinfo{author}{\bibfnamefont{K.~C.} \bibnamefont{Ku}},
  \bibinfo{author}{\bibfnamefont{S.~H.} \bibnamefont{Chun}},
  \bibinfo{author}{\bibfnamefont{J.~J.} \bibnamefont{Berry}},
  \bibinfo{author}{\bibfnamefont{N.}~\bibnamefont{Samarth}}, \bibnamefont{and}
  \bibinfo{author}{\bibfnamefont{P.}~\bibnamefont{Schiffer}},
  \bibinfo{journal}{Appl. Phys. Lett.} \textbf{\bibinfo{volume}{79}},
  \bibinfo{pages}{1495} (\bibinfo{year}{2001}).

\bibitem[{\citenamefont{Rushforth et~al.}(2008)\citenamefont{Rushforth,
  Ranieri, Zemen, Wunderlich, Edmonds, King, Ahmad, Campion, Foxon, Gallagher
  et~al.}}]{Rushforth:2008_a}
\bibinfo{author}{\bibfnamefont{A.~W.} \bibnamefont{Rushforth}},
  \bibinfo{author}{\bibfnamefont{E.~D.} \bibnamefont{Ranieri}},
  \bibinfo{author}{\bibfnamefont{J.}~\bibnamefont{Zemen}},
  \bibinfo{author}{\bibfnamefont{J.}~\bibnamefont{Wunderlich}},
  \bibinfo{author}{\bibfnamefont{K.~W.} \bibnamefont{Edmonds}},
  \bibinfo{author}{\bibfnamefont{C.~S.} \bibnamefont{King}},
  \bibinfo{author}{\bibfnamefont{E.}~\bibnamefont{Ahmad}},
  \bibinfo{author}{\bibfnamefont{R.~P.} \bibnamefont{Campion}},
  \bibinfo{author}{\bibfnamefont{C.~T.} \bibnamefont{Foxon}},
  \bibinfo{author}{\bibfnamefont{B.~L.} \bibnamefont{Gallagher}},
  \bibnamefont{et~al.}, \bibinfo{journal}{Phys. Rev.}
  \textbf{\bibinfo{volume}{B 78}}, \bibinfo{pages}{085314}
  (\bibinfo{year}{2008}), \eprint{arXiv:0801.0886}.

\bibitem[{\citenamefont{{de {Ranieri}} et~al.}(2008)\citenamefont{{de
  {Ranieri}}, Rushforth, {V\'{y}born\'{y}}, Rana, Ahmad, Campion, Foxon,
  Gallagher, Irvine, Wunderlich et~al.}}]{Ranieri:2008_a}
\bibinfo{author}{\bibfnamefont{E.}~\bibnamefont{{de {Ranieri}}}},
  \bibinfo{author}{\bibfnamefont{A.~W.} \bibnamefont{Rushforth}},
  \bibinfo{author}{\bibfnamefont{K.}~\bibnamefont{{V\'{y}born\'{y}}}},
  \bibinfo{author}{\bibfnamefont{U.}~\bibnamefont{Rana}},
  \bibinfo{author}{\bibfnamefont{E.}~\bibnamefont{Ahmad}},
  \bibinfo{author}{\bibfnamefont{R.~P.} \bibnamefont{Campion}},
  \bibinfo{author}{\bibfnamefont{C.~T.} \bibnamefont{Foxon}},
  \bibinfo{author}{\bibfnamefont{B.~L.} \bibnamefont{Gallagher}},
  \bibinfo{author}{\bibfnamefont{A.~C.} \bibnamefont{Irvine}},
  \bibinfo{author}{\bibfnamefont{J.}~\bibnamefont{Wunderlich}},
  \bibnamefont{et~al.}, \bibinfo{journal}{New J. Phys.}
  \textbf{\bibinfo{volume}{10}}, \bibinfo{pages}{065003}
  (\bibinfo{year}{2008}), \eprint{arXiv:0802.3344}.

\bibitem[{\citenamefont{Overby et~al.}(2008)\citenamefont{Overby, Chernyshov,
  Rokhinson, Liu, and Furdyna}}]{Overby:2008_a}
\bibinfo{author}{\bibfnamefont{M.}~\bibnamefont{Overby}},
  \bibinfo{author}{\bibfnamefont{A.}~\bibnamefont{Chernyshov}},
  \bibinfo{author}{\bibfnamefont{L.~P.} \bibnamefont{Rokhinson}},
  \bibinfo{author}{\bibfnamefont{X.}~\bibnamefont{Liu}}, \bibnamefont{and}
  \bibinfo{author}{\bibfnamefont{J.~K.} \bibnamefont{Furdyna}},
  \bibinfo{journal}{Appl. Phys. Lett.} \textbf{\bibinfo{volume}{92}},
  \bibinfo{pages}{192501} (\bibinfo{year}{2008}), \eprint{arXiv:0801.4191}.

\bibitem[{\citenamefont{Chiba et~al.}(2008)\citenamefont{Chiba, Sawicki,
  Nishitani, Nakatani, Matsukura, and Ohno}}]{Chiba:2008_a}
\bibinfo{author}{\bibfnamefont{D.}~\bibnamefont{Chiba}},
  \bibinfo{author}{\bibfnamefont{M.}~\bibnamefont{Sawicki}},
  \bibinfo{author}{\bibfnamefont{Y.}~\bibnamefont{Nishitani}},
  \bibinfo{author}{\bibfnamefont{Y.}~\bibnamefont{Nakatani}},
  \bibinfo{author}{\bibfnamefont{F.}~\bibnamefont{Matsukura}},
  \bibnamefont{and} \bibinfo{author}{\bibfnamefont{H.}~\bibnamefont{Ohno}},
  \bibinfo{journal}{Nature} \textbf{\bibinfo{volume}{455}},
  \bibinfo{pages}{515} (\bibinfo{year}{2008}).

\bibitem[{\citenamefont{Owen et~al.}(2009)\citenamefont{Owen, Wunderlich,
  {Nov\'{a}k}, {Olejn\'{i}k}, Zemen, {V\'{y}born\'{y}}, Ogawa, Irvine,
  Ferguson, Sirringhaus et~al.}}]{Owen:2008_a}
\bibinfo{author}{\bibfnamefont{M.~H.~S.} \bibnamefont{Owen}},
  \bibinfo{author}{\bibfnamefont{J.}~\bibnamefont{Wunderlich}},
  \bibinfo{author}{\bibfnamefont{V.}~\bibnamefont{{Nov\'{a}k}}},
  \bibinfo{author}{\bibfnamefont{K.}~\bibnamefont{{Olejn\'{i}k}}},
  \bibinfo{author}{\bibfnamefont{J.}~\bibnamefont{Zemen}},
  \bibinfo{author}{\bibfnamefont{K.}~\bibnamefont{{V\'{y}born\'{y}}}},
  \bibinfo{author}{\bibfnamefont{S.}~\bibnamefont{Ogawa}},
  \bibinfo{author}{\bibfnamefont{A.~C.} \bibnamefont{Irvine}},
  \bibinfo{author}{\bibfnamefont{A.~J.} \bibnamefont{Ferguson}},
  \bibinfo{author}{\bibfnamefont{H.}~\bibnamefont{Sirringhaus}},
  \bibnamefont{et~al.}, \bibinfo{journal}{New J. Phys.}
  \textbf{\bibinfo{volume}{11}} (\bibinfo{year}{2009}),
  \eprint{arXiv:0807.0906}.

\bibitem[{\citenamefont{Wunderlich et~al.}(2007)\citenamefont{Wunderlich,
  Irvine, Zemen, {Hol\'{y}}, Rushforth, Ranieri, Rana, {V\'{y}born\'{y}},
  Sinova, Foxon et~al.}}]{Wunderlich:2007_c}
\bibinfo{author}{\bibfnamefont{J.}~\bibnamefont{Wunderlich}},
  \bibinfo{author}{\bibfnamefont{A.~C.} \bibnamefont{Irvine}},
  \bibinfo{author}{\bibfnamefont{J.}~\bibnamefont{Zemen}},
  \bibinfo{author}{\bibfnamefont{V.}~\bibnamefont{{Hol\'{y}}}},
  \bibinfo{author}{\bibfnamefont{A.~W.} \bibnamefont{Rushforth}},
  \bibinfo{author}{\bibfnamefont{E.~D.} \bibnamefont{Ranieri}},
  \bibinfo{author}{\bibfnamefont{U.}~\bibnamefont{Rana}},
  \bibinfo{author}{\bibfnamefont{K.}~\bibnamefont{{V\'{y}born\'{y}}}},
  \bibinfo{author}{\bibfnamefont{J.}~\bibnamefont{Sinova}},
  \bibinfo{author}{\bibfnamefont{C.~T.} \bibnamefont{Foxon}},
  \bibnamefont{et~al.}, \bibinfo{journal}{Phys. Rev. B}
  \textbf{\bibinfo{volume}{76}}, \bibinfo{pages}{054424}
  (\bibinfo{year}{2007}), \eprint{arXiv:0707.3329}.

\bibitem[{\citenamefont{Pappert et~al.}(2007)\citenamefont{Pappert,
  {H\"{u}mpfner}, Gould, Wenisch, Brunner, Schmidt, and
  Molenkamp}}]{Pappert:2007_a}
\bibinfo{author}{\bibfnamefont{K.}~\bibnamefont{Pappert}},
  \bibinfo{author}{\bibfnamefont{S.}~\bibnamefont{{H\"{u}mpfner}}},
  \bibinfo{author}{\bibfnamefont{C.}~\bibnamefont{Gould}},
  \bibinfo{author}{\bibfnamefont{J.}~\bibnamefont{Wenisch}},
  \bibinfo{author}{\bibfnamefont{K.}~\bibnamefont{Brunner}},
  \bibinfo{author}{\bibfnamefont{G.}~\bibnamefont{Schmidt}}, \bibnamefont{and}
  \bibinfo{author}{\bibfnamefont{L.~W.} \bibnamefont{Molenkamp}},
  \bibinfo{journal}{Nature Phys.} \textbf{\bibinfo{volume}{3}},
  \bibinfo{pages}{573} (\bibinfo{year}{2007}), \eprint{arXiv:cond-mat/0701478}.

\bibitem[{\citenamefont{Wenisch et~al.}(2007)\citenamefont{Wenisch, Gould,
  Ebel, Storz, Pappert, Schmidt, Kumpf, Schmidt, Brunner, and
  Molenkamp}}]{Wenisch:2007_a}
\bibinfo{author}{\bibfnamefont{J.}~\bibnamefont{Wenisch}},
  \bibinfo{author}{\bibfnamefont{C.}~\bibnamefont{Gould}},
  \bibinfo{author}{\bibfnamefont{L.}~\bibnamefont{Ebel}},
  \bibinfo{author}{\bibfnamefont{J.}~\bibnamefont{Storz}},
  \bibinfo{author}{\bibfnamefont{K.}~\bibnamefont{Pappert}},
  \bibinfo{author}{\bibfnamefont{M.~J.} \bibnamefont{Schmidt}},
  \bibinfo{author}{\bibfnamefont{C.}~\bibnamefont{Kumpf}},
  \bibinfo{author}{\bibfnamefont{G.}~\bibnamefont{Schmidt}},
  \bibinfo{author}{\bibfnamefont{K.}~\bibnamefont{Brunner}}, \bibnamefont{and}
  \bibinfo{author}{\bibfnamefont{L.~W.} \bibnamefont{Molenkamp}},
  \bibinfo{journal}{Phys. Rev. Lett.} \textbf{\bibinfo{volume}{99}},
  \bibinfo{pages}{077201} (\bibinfo{year}{2007}),
  \eprint{arXiv:cond-mat/0701479}.

\bibitem[{\citenamefont{{H\"{u}mpfner}
  et~al.}(2007)\citenamefont{{H\"{u}mpfner}, Sawicki, Pappert, Wenisch,
  Brunner, Gould, Schmidt, Dietl, and Molenkamp}}]{Humpfner:2006_a}
\bibinfo{author}{\bibfnamefont{S.}~\bibnamefont{{H\"{u}mpfner}}},
  \bibinfo{author}{\bibfnamefont{M.}~\bibnamefont{Sawicki}},
  \bibinfo{author}{\bibfnamefont{K.}~\bibnamefont{Pappert}},
  \bibinfo{author}{\bibfnamefont{J.}~\bibnamefont{Wenisch}},
  \bibinfo{author}{\bibfnamefont{K.}~\bibnamefont{Brunner}},
  \bibinfo{author}{\bibfnamefont{C.}~\bibnamefont{Gould}},
  \bibinfo{author}{\bibfnamefont{G.}~\bibnamefont{Schmidt}},
  \bibinfo{author}{\bibfnamefont{T.}~\bibnamefont{Dietl}}, \bibnamefont{and}
  \bibinfo{author}{\bibfnamefont{L.~W.} \bibnamefont{Molenkamp}},
  \bibinfo{journal}{Appl. Phys. Lett.} \textbf{\bibinfo{volume}{90}},
  \bibinfo{pages}{102102} (\bibinfo{year}{2007}),
  \eprint{arXiv:cond-mat/0612439}.

\bibitem[{\citenamefont{Rushforth et~al.}(2007)\citenamefont{Rushforth,
  {V\'{y}born\'{y}}, King, Edmonds, Campion, Foxon, Wunderlich, Irvine,
  {Va\v{s}ek}, {Nov\'{a}k} et~al.}}]{Rushforth:2007_a}
\bibinfo{author}{\bibfnamefont{A.~W.} \bibnamefont{Rushforth}},
  \bibinfo{author}{\bibfnamefont{K.}~\bibnamefont{{V\'{y}born\'{y}}}},
  \bibinfo{author}{\bibfnamefont{C.~S.} \bibnamefont{King}},
  \bibinfo{author}{\bibfnamefont{K.~W.} \bibnamefont{Edmonds}},
  \bibinfo{author}{\bibfnamefont{R.~P.} \bibnamefont{Campion}},
  \bibinfo{author}{\bibfnamefont{C.~T.} \bibnamefont{Foxon}},
  \bibinfo{author}{\bibfnamefont{J.}~\bibnamefont{Wunderlich}},
  \bibinfo{author}{\bibfnamefont{A.~C.} \bibnamefont{Irvine}},
  \bibinfo{author}{\bibfnamefont{P.}~\bibnamefont{{Va\v{s}ek}}},
  \bibinfo{author}{\bibfnamefont{V.}~\bibnamefont{{Nov\'{a}k}}},
  \bibnamefont{et~al.}, \bibinfo{journal}{Phys. Rev. Lett.}
  \textbf{\bibinfo{volume}{99}}, \bibinfo{pages}{147207}
  (\bibinfo{year}{2007}), \eprint{arXiv:cond-mat/0702357}.

\bibitem[{\citenamefont{Dietl et~al.}(2001)\citenamefont{Dietl, Ohno, and
  Matsukura}}]{Dietl:2001_b}
\bibinfo{author}{\bibfnamefont{T.}~\bibnamefont{Dietl}},
  \bibinfo{author}{\bibfnamefont{H.}~\bibnamefont{Ohno}}, \bibnamefont{and}
  \bibinfo{author}{\bibfnamefont{F.}~\bibnamefont{Matsukura}},
  \bibinfo{journal}{Phys. Rev.} \textbf{\bibinfo{volume}{B 63}},
  \bibinfo{pages}{195205} (\bibinfo{year}{2001}),
  \eprint{arXiv:cond-mat/0007190}.

\bibitem[{\citenamefont{Abolfath et~al.}(2001)\citenamefont{Abolfath,
  Jungwirth, Brum, and MacDonald}}]{Abolfath:2001_a}
\bibinfo{author}{\bibfnamefont{M.}~\bibnamefont{Abolfath}},
  \bibinfo{author}{\bibfnamefont{T.}~\bibnamefont{Jungwirth}},
  \bibinfo{author}{\bibfnamefont{J.}~\bibnamefont{Brum}}, \bibnamefont{and}
  \bibinfo{author}{\bibfnamefont{A.~H.} \bibnamefont{MacDonald}},
  \bibinfo{journal}{Phys. Rev.} \textbf{\bibinfo{volume}{B 63}},
  \bibinfo{pages}{054418} (\bibinfo{year}{2001}),
  \eprint{arXiv:cond-mat/0006093}.

\bibitem[{\citenamefont{Jungwirth et~al.}(2006)\citenamefont{Jungwirth, Sinova,
  {Ma\v{s}ek}, {Ku\v{c}era}, and MacDonald}}]{Jungwirth:2006_a}
\bibinfo{author}{\bibfnamefont{T.}~\bibnamefont{Jungwirth}},
  \bibinfo{author}{\bibfnamefont{J.}~\bibnamefont{Sinova}},
  \bibinfo{author}{\bibfnamefont{J.}~\bibnamefont{{Ma\v{s}ek}}},
  \bibinfo{author}{\bibfnamefont{J.}~\bibnamefont{{Ku\v{c}era}}},
  \bibnamefont{and} \bibinfo{author}{\bibfnamefont{A.~H.}
  \bibnamefont{MacDonald}}, \bibinfo{journal}{Rev. Mod. Phys.}
  \textbf{\bibinfo{volume}{78}}, \bibinfo{pages}{809} (\bibinfo{year}{2006}),
  \eprint{arXiv:cond-mat/0603380}.

\bibitem[{\citenamefont{Jungwirth
  et~al.}(2005{\natexlab{a}})\citenamefont{Jungwirth, Wang, {Ma\v{s}ek},
  Edmonds, {K\"{o}nig}, Sinova, Polini, Goncharuk, MacDonald, Sawicki
  et~al.}}]{Jungwirth:2005_b}
\bibinfo{author}{\bibfnamefont{T.}~\bibnamefont{Jungwirth}},
  \bibinfo{author}{\bibfnamefont{K.~Y.} \bibnamefont{Wang}},
  \bibinfo{author}{\bibfnamefont{J.}~\bibnamefont{{Ma\v{s}ek}}},
  \bibinfo{author}{\bibfnamefont{K.~W.} \bibnamefont{Edmonds}},
  \bibinfo{author}{\bibfnamefont{J.}~\bibnamefont{{K\"{o}nig}}},
  \bibinfo{author}{\bibfnamefont{J.}~\bibnamefont{Sinova}},
  \bibinfo{author}{\bibfnamefont{M.}~\bibnamefont{Polini}},
  \bibinfo{author}{\bibfnamefont{N.~A.} \bibnamefont{Goncharuk}},
  \bibinfo{author}{\bibfnamefont{A.~H.} \bibnamefont{MacDonald}},
  \bibinfo{author}{\bibfnamefont{M.}~\bibnamefont{Sawicki}},
  \bibnamefont{et~al.}, \bibinfo{journal}{Phys. Rev.}
  \textbf{\bibinfo{volume}{B 72}}, \bibinfo{pages}{165204}
  (\bibinfo{year}{2005}{\natexlab{a}}), \eprint{arXiv:cond-mat/0505215}.

\bibitem[{\citenamefont{{Ma\v{s}ek} et~al.}(2003)\citenamefont{{Ma\v{s}ek},
  {Kudrnovsk\'{y}}, and {M\'{a}ca}}}]{Masek:2003_a}
\bibinfo{author}{\bibfnamefont{J.}~\bibnamefont{{Ma\v{s}ek}}},
  \bibinfo{author}{\bibfnamefont{J.}~\bibnamefont{{Kudrnovsk\'{y}}}},
  \bibnamefont{and}
  \bibinfo{author}{\bibfnamefont{F.}~\bibnamefont{{M\'{a}ca}}},
  \bibinfo{journal}{Phys. Rev.} \textbf{\bibinfo{volume}{B 67}},
  \bibinfo{pages}{153203} (\bibinfo{year}{2003}),
  \eprint{arXiv:cond-mat/0302150}.

\bibitem[{\citenamefont{Daeubler et~al.}(2008)\citenamefont{Daeubler,
  Schwaiger, Glunk, Tabor, Schoch, Sauer, and Limmer}}]{Daeubler:2007_a}
\bibinfo{author}{\bibfnamefont{J.}~\bibnamefont{Daeubler}},
  \bibinfo{author}{\bibfnamefont{S.}~\bibnamefont{Schwaiger}},
  \bibinfo{author}{\bibfnamefont{M.}~\bibnamefont{Glunk}},
  \bibinfo{author}{\bibfnamefont{M.}~\bibnamefont{Tabor}},
  \bibinfo{author}{\bibfnamefont{W.}~\bibnamefont{Schoch}},
  \bibinfo{author}{\bibfnamefont{R.}~\bibnamefont{Sauer}}, \bibnamefont{and}
  \bibinfo{author}{\bibfnamefont{W.}~\bibnamefont{Limmer}},
  \bibinfo{journal}{Physica} p. \bibinfo{pages}{1876} (\bibinfo{year}{2008}).

\bibitem[{\citenamefont{Zemen et~al.}(2009)\citenamefont{Zemen, Kucera,
  Olejnik, and Jungwirth}}]{Zemen:2009_a}
\bibinfo{author}{\bibfnamefont{J.}~\bibnamefont{Zemen}},
  \bibinfo{author}{\bibfnamefont{J.}~\bibnamefont{Kucera}},
  \bibinfo{author}{\bibfnamefont{K.}~\bibnamefont{Olejnik}}, \bibnamefont{and}
  \bibinfo{author}{\bibfnamefont{T.}~\bibnamefont{Jungwirth}},
  \bibinfo{journal}{Phys. Rev.} \textbf{\bibinfo{volume}{B 80}},
  \bibinfo{pages}{155203} (\bibinfo{year}{2009}), \eprint{arXiv:0904.0993}.

\bibitem[{\citenamefont{Zhao et~al.}(2005)\citenamefont{Zhao, Staddon, Wang,
  Edmonds, Campion, Gallagher, and Foxon}}]{Zhao:2005_a}
\bibinfo{author}{\bibfnamefont{L.~X.} \bibnamefont{Zhao}},
  \bibinfo{author}{\bibfnamefont{C.~R.} \bibnamefont{Staddon}},
  \bibinfo{author}{\bibfnamefont{K.~Y.} \bibnamefont{Wang}},
  \bibinfo{author}{\bibfnamefont{K.~W.} \bibnamefont{Edmonds}},
  \bibinfo{author}{\bibfnamefont{R.~P.} \bibnamefont{Campion}},
  \bibinfo{author}{\bibfnamefont{B.~L.} \bibnamefont{Gallagher}},
  \bibnamefont{and} \bibinfo{author}{\bibfnamefont{C.~T.} \bibnamefont{Foxon}},
  \bibinfo{journal}{Appl. Phys. Lett.} \textbf{\bibinfo{volume}{86}},
  \bibinfo{pages}{071902} (\bibinfo{year}{2005}),
  \eprint{arXiv:cond-mat/0501314}.

\bibitem[{\citenamefont{Liu et~al.}(2005)\citenamefont{Liu, Lim, Titova,
  Dobrowolska, Furdyna, Kutrowski, and Wojtowicz}}]{Liu:2005_d}
\bibinfo{author}{\bibfnamefont{X.}~\bibnamefont{Liu}},
  \bibinfo{author}{\bibfnamefont{W.~L.} \bibnamefont{Lim}},
  \bibinfo{author}{\bibfnamefont{L.~V.} \bibnamefont{Titova}},
  \bibinfo{author}{\bibfnamefont{M.}~\bibnamefont{Dobrowolska}},
  \bibinfo{author}{\bibfnamefont{J.~K.} \bibnamefont{Furdyna}},
  \bibinfo{author}{\bibfnamefont{M.}~\bibnamefont{Kutrowski}},
  \bibnamefont{and}
  \bibinfo{author}{\bibfnamefont{T.}~\bibnamefont{Wojtowicz}},
  \bibinfo{journal}{J. Appl. Phys.} \textbf{\bibinfo{volume}{98}},
  \bibinfo{pages}{063904} (\bibinfo{year}{2005}),
  \eprint{arXiv:cond-mat/0505322}.

\bibitem[{\citenamefont{Welp et~al.}(2003)\citenamefont{Welp, Vlasko-Vlasov,
  Liu, Furdyna, and Wojtowicz}}]{Welp:2003_a}
\bibinfo{author}{\bibfnamefont{U.}~\bibnamefont{Welp}},
  \bibinfo{author}{\bibfnamefont{V.~K.} \bibnamefont{Vlasko-Vlasov}},
  \bibinfo{author}{\bibfnamefont{X.}~\bibnamefont{Liu}},
  \bibinfo{author}{\bibfnamefont{J.~K.} \bibnamefont{Furdyna}},
  \bibnamefont{and}
  \bibinfo{author}{\bibfnamefont{T.}~\bibnamefont{Wojtowicz}},
  \bibinfo{journal}{Phys. Rev. Lett.} \textbf{\bibinfo{volume}{90}},
  \bibinfo{pages}{167206} (\bibinfo{year}{2003}).

\bibitem[{\citenamefont{Thevenard et~al.}(2007)\citenamefont{Thevenard,
  Largeau, Mauguin, {Lema\^{i}tre}, Khazen, and von
  Bardeleben}}]{Thevenard:2007_a}
\bibinfo{author}{\bibfnamefont{L.}~\bibnamefont{Thevenard}},
  \bibinfo{author}{\bibfnamefont{L.}~\bibnamefont{Largeau}},
  \bibinfo{author}{\bibfnamefont{O.}~\bibnamefont{Mauguin}},
  \bibinfo{author}{\bibfnamefont{A.}~\bibnamefont{{Lema\^{i}tre}}},
  \bibinfo{author}{\bibfnamefont{K.}~\bibnamefont{Khazen}}, \bibnamefont{and}
  \bibinfo{author}{\bibfnamefont{H.~J.} \bibnamefont{von Bardeleben}},
  \bibinfo{journal}{Phys. Rev.} \textbf{\bibinfo{volume}{B 75}},
  \bibinfo{pages}{195218} (\bibinfo{year}{2007}),
  \eprint{arXiv:cond-mat/0702548}.

\bibitem[{\citenamefont{Aharoni}(1998)}]{Aharoni:1997_a}
\bibinfo{author}{\bibfnamefont{A.}~\bibnamefont{Aharoni}},
  \bibinfo{journal}{Journal of Applied Physics} p. \bibinfo{pages}{3432}
  (\bibinfo{year}{1998}).

\bibitem[{\citenamefont{Jungwirth
  et~al.}(2005{\natexlab{b}})\citenamefont{Jungwirth, {Ma\v{s}ek}, Wang,
  Edmonds, Sawicki, Polini, Sinova, MacDonald, Campion, Zhao
  et~al.}}]{Jungwirth:2005_a}
\bibinfo{author}{\bibfnamefont{T.}~\bibnamefont{Jungwirth}},
  \bibinfo{author}{\bibfnamefont{J.}~\bibnamefont{{Ma\v{s}ek}}},
  \bibinfo{author}{\bibfnamefont{K.~Y.} \bibnamefont{Wang}},
  \bibinfo{author}{\bibfnamefont{K.~W.} \bibnamefont{Edmonds}},
  \bibinfo{author}{\bibfnamefont{M.}~\bibnamefont{Sawicki}},
  \bibinfo{author}{\bibfnamefont{M.}~\bibnamefont{Polini}},
  \bibinfo{author}{\bibfnamefont{J.}~\bibnamefont{Sinova}},
  \bibinfo{author}{\bibfnamefont{A.~H.} \bibnamefont{MacDonald}},
  \bibinfo{author}{\bibfnamefont{R.~P.} \bibnamefont{Campion}},
  \bibinfo{author}{\bibfnamefont{L.~X.} \bibnamefont{Zhao}},
  \bibnamefont{et~al.}, \bibinfo{journal}{Phys. Rev.}
  \textbf{\bibinfo{volume}{B 73}}, \bibinfo{pages}{165205}
  (\bibinfo{year}{2005}{\natexlab{b}}), \eprint{arXiv:cond-mat/0508255}.

\bibitem[{\citenamefont{Sawicki et~al.}(2005)\citenamefont{Sawicki, Wang,
  Edmonds, Campion, Staddon, Farley, Foxon, Papis, Kaminska, Piotrowska
  et~al.}}]{Sawicki:2004_a}
\bibinfo{author}{\bibfnamefont{M.}~\bibnamefont{Sawicki}},
  \bibinfo{author}{\bibfnamefont{K.-Y.} \bibnamefont{Wang}},
  \bibinfo{author}{\bibfnamefont{K.~W.} \bibnamefont{Edmonds}},
  \bibinfo{author}{\bibfnamefont{R.~P.} \bibnamefont{Campion}},
  \bibinfo{author}{\bibfnamefont{C.~R.} \bibnamefont{Staddon}},
  \bibinfo{author}{\bibfnamefont{N.~R.~S.} \bibnamefont{Farley}},
  \bibinfo{author}{\bibfnamefont{C.~T.} \bibnamefont{Foxon}},
  \bibinfo{author}{\bibfnamefont{E.}~\bibnamefont{Papis}},
  \bibinfo{author}{\bibfnamefont{E.}~\bibnamefont{Kaminska}},
  \bibinfo{author}{\bibfnamefont{A.}~\bibnamefont{Piotrowska}},
  \bibnamefont{et~al.}, \bibinfo{journal}{Phys. Rev.}
  \textbf{\bibinfo{volume}{B 71}}, \bibinfo{pages}{121302}
  (\bibinfo{year}{2005}), \eprint{arXiv:cond-mat/0410544}.

\end{thebibliography}
\end{document}